\documentclass{emulateapj} 
\usepackage{psfig}
\usepackage{mathrsfs}
\usepackage{rotating}
\begin{document}
\slugcomment{ApJ Supplement Series, in press}
\newcommand{\LeeZinn}{\mathscr{L}}

\shortauthors{M. Catelan, B. J. Pritzl, H. A. Smith} 
\shorttitle{The RRL PL relation in $UBVRIJHK$}

\title{The RR Lyrae Period-Luminosity Relation. \\ I. Theoretical Calibration}

\author{M.~Catelan} 
\affil{Pontificia Universidad Cat\'olica de Chile, Departamento de 
       Astronom\'\i a y Astrof\'\i sica, \\ Av. Vicu\~{n}a Mackenna 4860, 
      782-0436 Macul, Santiago, Chile}
\email{mcatelan@astro.puc.cl} 

\author{Barton J. Pritzl}  
\affil{Macalester College, 1600 Grand Avenue, Saint Paul, MN 55105} 
\email{pritzl@macalester.edu}

\and
\author{Horace A. Smith}  
\affil{Dept.\ of Physics and Astronomy, Michigan State University, 
       East Lansing, MI 48824} 
\email{smith@pa.msu.edu}

\begin{abstract}
We present a theoretical calibration of the RR Lyrae period-luminosity (PL) 
relation in the $UBVRIJHK$ Johnsons-Cousins-Glass system. Our theoretical 
work is based on calculations of synthetic horizontal branches (HBs) for 
several different metallicities, fully taking into account evolutionary 
effects besides the effect of chemical composition. Extensive tabulations 
of our results are provided, including convenient analytical formulae for 
the calculation of the coefficients of the period-luminosity relation in 
the different passbands as a function of HB type. 
We also provide ``average'' PL relations in $IJHK$, 
for applications in cases where the HB type is not known a priori; as well 
as a new calibration of the $M_V - {\rm [M/H]}$ relation. These can be 
summarized as follows:  

\begin{displaymath}
M_I = 0.471  - 1.132 \, \log P + 0.205 \, \log Z, 
\end{displaymath} 

\begin{displaymath}
M_J = -0.141  - 1.773 \, \log P + 0.190 \, \log Z, 
\end{displaymath} 

\begin{displaymath}
M_H = -0.551 - 2.313 \, \log P + 0.178 \, \log Z, 
\end{displaymath} 
 
\begin{displaymath}
M_K = -0.597 - 2.353 \, \log P + 0.175 \, \log Z,   
\end{displaymath} 

\noindent and 

\begin{displaymath}
M_V = 2.288 + 0.882 \, \log Z + 0.108 \, (\log Z)^{2}.  
\end{displaymath}

\end{abstract}


\section{Introduction}
RR Lyrae (RRL) stars are the cornerstone of the Population~II distance 
scale. Yet, unlike Cepheids, which have for almost a century been known 
to present a tight period-luminosity (PL) relation (Leavitt 1912), RRL 
have not been known for presenting a particularly noteworthy PL relation. 
Instead, most researchers have utilized an average relation between 
absolute visual magnitude and metallicity [Fe/H] when deriving RRL-based 
distances. This relation possesses several potential pitfalls, including 
a strong dependence on evolutionary effects (e.g., Demarque et al. 2000), 
a possible non-linearity as a function of [Fe/H] (e.g., Castellani, Chieffi,  
\& Pulone 1991), and ``pathological outliers'' (e.g., Pritzl et al. 2002).

\begin{figure*}[t]
  \figurenum{1}
  \epsscale{1}
\plotone{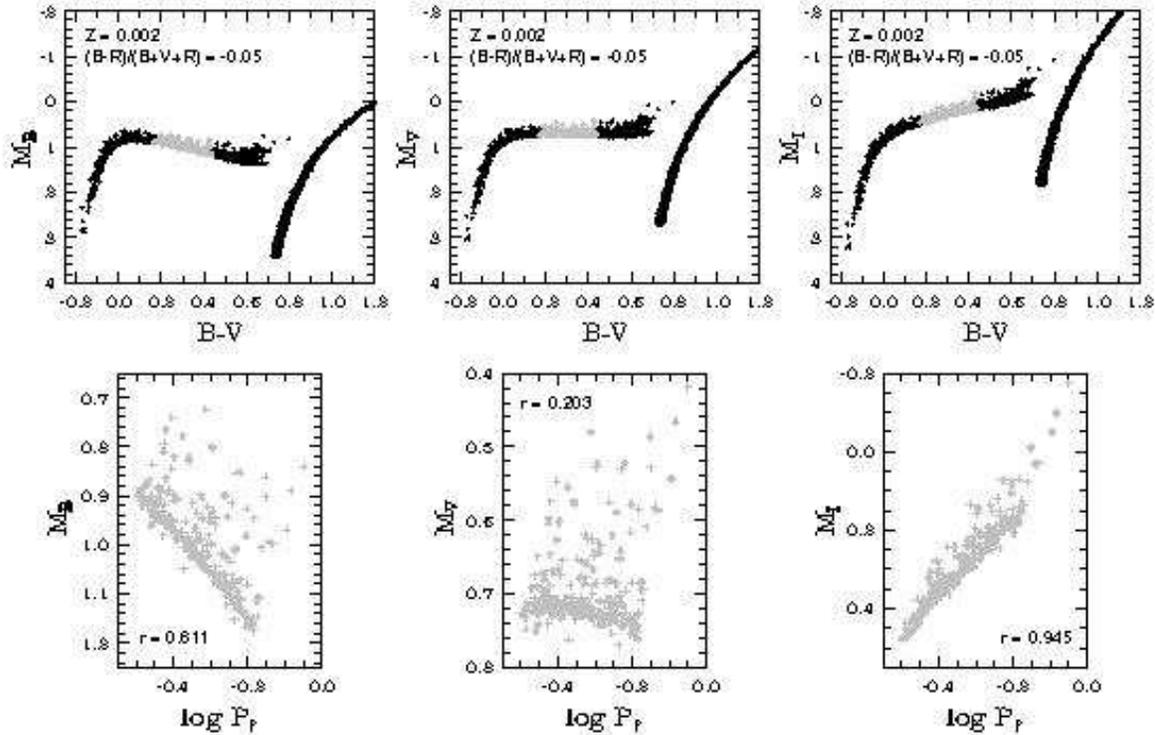}
  \caption{{\em Upper panels}: Morphology of the HB in different bandpasses 
     (left: $B$; middle: $V$; right: $I$). RRL variables are shown in gray, 
     and non-variable stars in black. {\em Lower panels}: Corresponding RRL
     distributions in the absolute magnitude---log-period plane. The 
     correlation coefficient $r$ is shown in the lower panels. All plots 
     refer to an HB simulation with $Z = 0.002$ and an intermediate HB type, 
     as indicated in the upper panels.  
      }
      \label{Fig01}
\end{figure*}

To be sure, RRL have also been noted to follow a PL relation, but only 
in the $K$ band (Longmore, Fernley, \& Jameson 1986). This is in sharp 
contrast with the case of the Cepheids, which follow tight PL relations 
both in the visual and in the near-infrared (see, e.g., Tanvir 1999). 
The reason why Cepheids present a tight PL relation irrespective of 
bandpass is that these stars cover a large range in luminosities but 
only a modest range in temperatures. Conversely, RRL stars are 
restricted to the horizontal branch (HB) phase of low-mass stars, and 
thus necessarily cover a much more modest range in luminosities---so 
much so that, in their case, the range in temperature of the instability 
strip is as important as, if not more important than, the range in 
luminosities of RRL stars, in determining their range in periods. 
Therefore, RRL stars may indeed present PL relations, but only if the 
bolometric corrections are such as to lead to a large range in absolute 
magnitudes when going from the blue to the red sides of the instability 
strip---as is indeed the case in $K$.

The purpose of the present paper, then, is to perform the first systematic 
analysis of whether a useful RRL PL relation may also be present in other 
bandpasses besides $K$. In particular, we expect that, using bandpasses 
in which the HB is not quite ``horizontal'' at the RRL level, a PL 
relation should indeed be present. Since the HB around the RRL region 
becomes distinctly non-horizontal both towards the near-ultraviolet (e.g., 
Fig.~4 in Ferraro et al. 1998) and towards the near-infrared (e.g., 
Davidge \& Courteau 1999), we present a full analysis of the slope and 
zero point of the RRL PL relation in the Johnsons-Cousins-Glass system, 
from $U$ to $K$, including also $BVRIJH$.

\section{Models}
The HB simulations employed in the present paper are similar to those 
described in Catelan (2004a), to which the reader is referred for further 
details and references about the HB synthesis method. The evolutionary tracks 
employed here are those computed by Catelan et al. (1998) for $Z = 0.001$ 
and $Z = 0.0005$, and by Sweigart \& Catelan (1998) for $Z = 0.002$ and 
$Z = 0.006$, and assume a main-sequence helium abundance of 23\% by 
mass and scaled-solar compositions. The 
mass distribution is represented by a normal deviate with a mass 
dispersion $\sigma_M = 0.020 \, M_{\odot}$. 
For the purposes of the present paper, we have added to this 
code bolometric corrections from Girardi et al. (2002) for $URJHK$ over 
the relevant ranges of temperature and gravity. The width of the 
instability strip is taken as $\Delta\log T_{\rm eff} = 0.075$, which
provides the temperature of the red edge of the instability strip for 
each star once its blue edge has been computed on the basis of RRL 
pulsation theory results. More specifically, the instability strip blue 
edge adopted in this paper is based on equation~(1) of Caputo et al. (1987), 
which provides a fit to Stellingwerf's (1984) results---except 
that a shift by $-200$~K to the temperature values thus derived was applied 
in order to improve agreement with more recent theoretical prescriptions  
(see \S6 in Catelan 2004a for a detailed discussion). 
We include both fundamental-mode (RRab) and ``fundamentalized'' 
first-overtone (RRc) variables in our final PL relations. The computed 
periods are based on equation~(4) in Caputo, Marconi, \& Santolamazza (1998), 
which represents an updated version of the van Albada \& Baker (1971) 
period-mean density relation.

\begin{figure*}[t]
  \figurenum{2}
  \epsscale{0.875}
\plotone{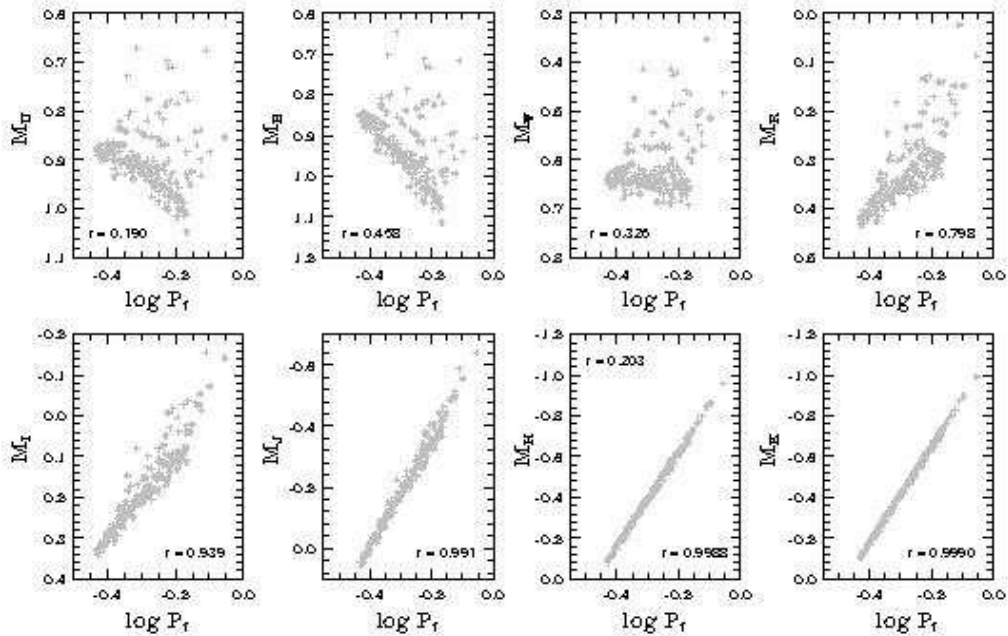}
  \caption{PL relations in several different passbands. {\em Upper panels}: 
    $U$ (left), $B$, $V$, $R$ (right). {\em Lower panels}: $I$ (left), $J$, 
    $H$, $K$ (right). The correlation coefficient is shown in all panels. 
    All plots refer to an HB simulation with $Z = 0.001$ and an intermediate 
    HB type.        }
      \label{Fig02}
\end{figure*}

\begin{figure*}[t]
  \figurenum{3}
  \epsscale{0.95}
  \plotone{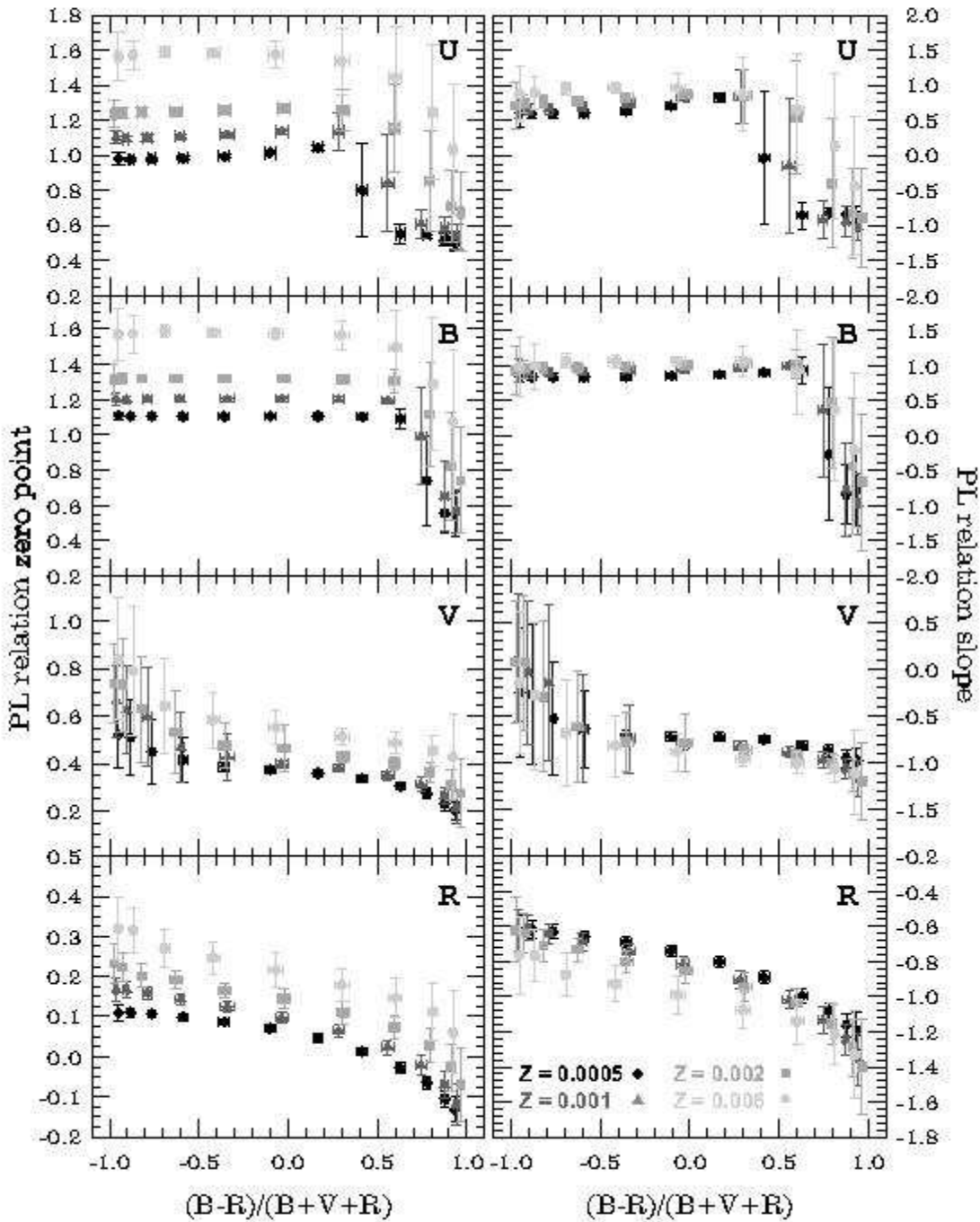}
  \caption{Theoretically calibrated PL relations in the $UBVR$ passbands
    (from top to bottom), for the four indicated metallicities. The zero 
    points (left panels) and slopes (right panels) are given as a function 
    of the Lee-Zinn HB morphology indicator. 
        }
      \label{Fig3}
\end{figure*}

\begin{figure*}[t]
  \figurenum{4}
  \epsscale{0.95}
  \plotone{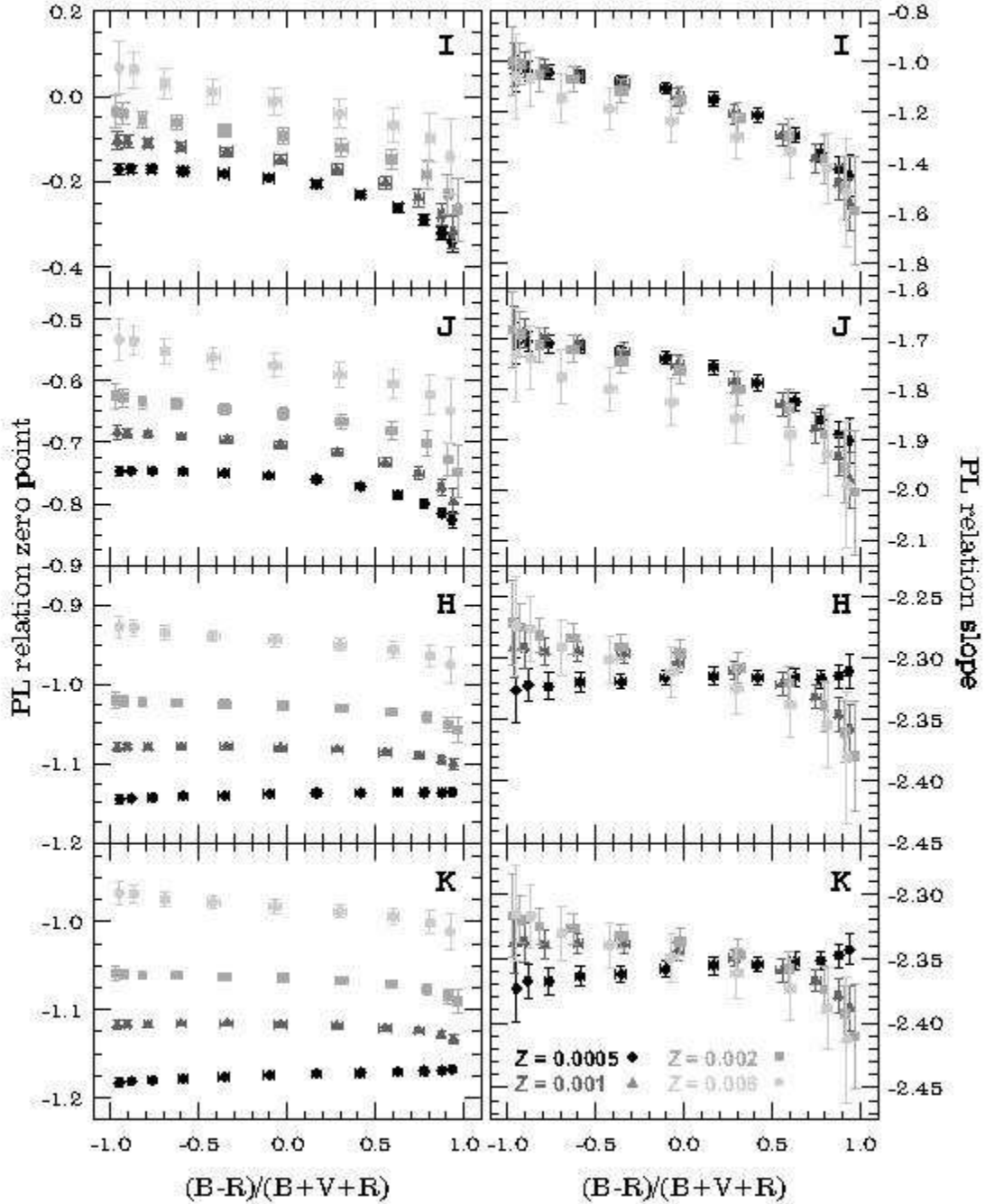}
  \caption{As in Figure~3, but for $IJHK$ (from top to bottom).  
        }
      \label{Fig4}
\end{figure*}

\begin{sidewaysfigure*}
  \figurenum{5}
  \epsscale{0.875}
  \plotone{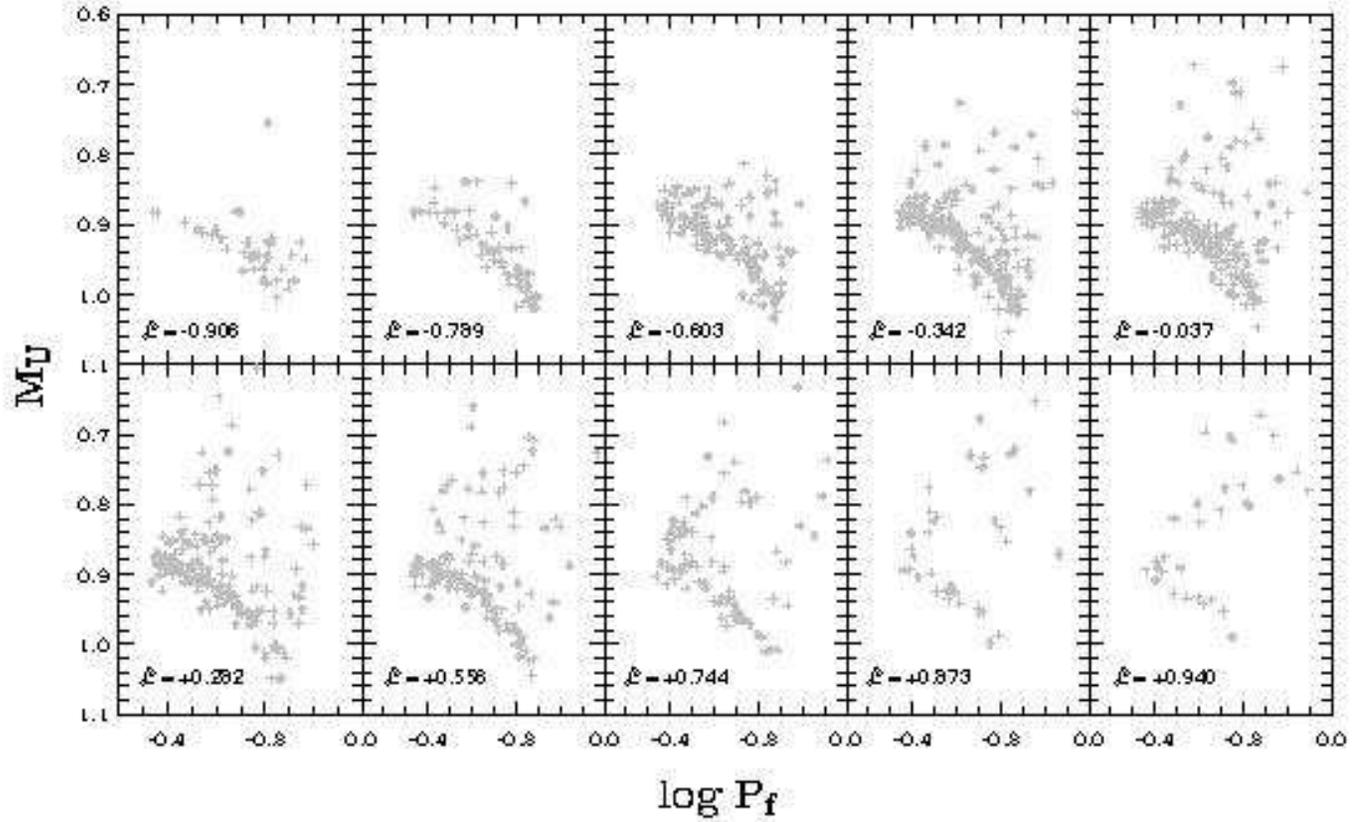}
  \caption{Variation in the $M_U - \log P$ relation as a function of 
     HB type, for a metallicity $Z = 0.001$. 
     The HB morphology, indicated by the $\LeeZinn$ value, 
     becomes bluer from upper left to lower right. For each HB type, 
     only the first in the series of 100 simulations used to compute 
     the average coefficients shown in Figures~3 and 4 and Table~1 was 
     chosen to produce this figure.    
        }
      \label{Fig5}
\end{sidewaysfigure*}

\begin{sidewaysfigure*}
  \figurenum{6}
  \epsscale{0.875}
  \plotone{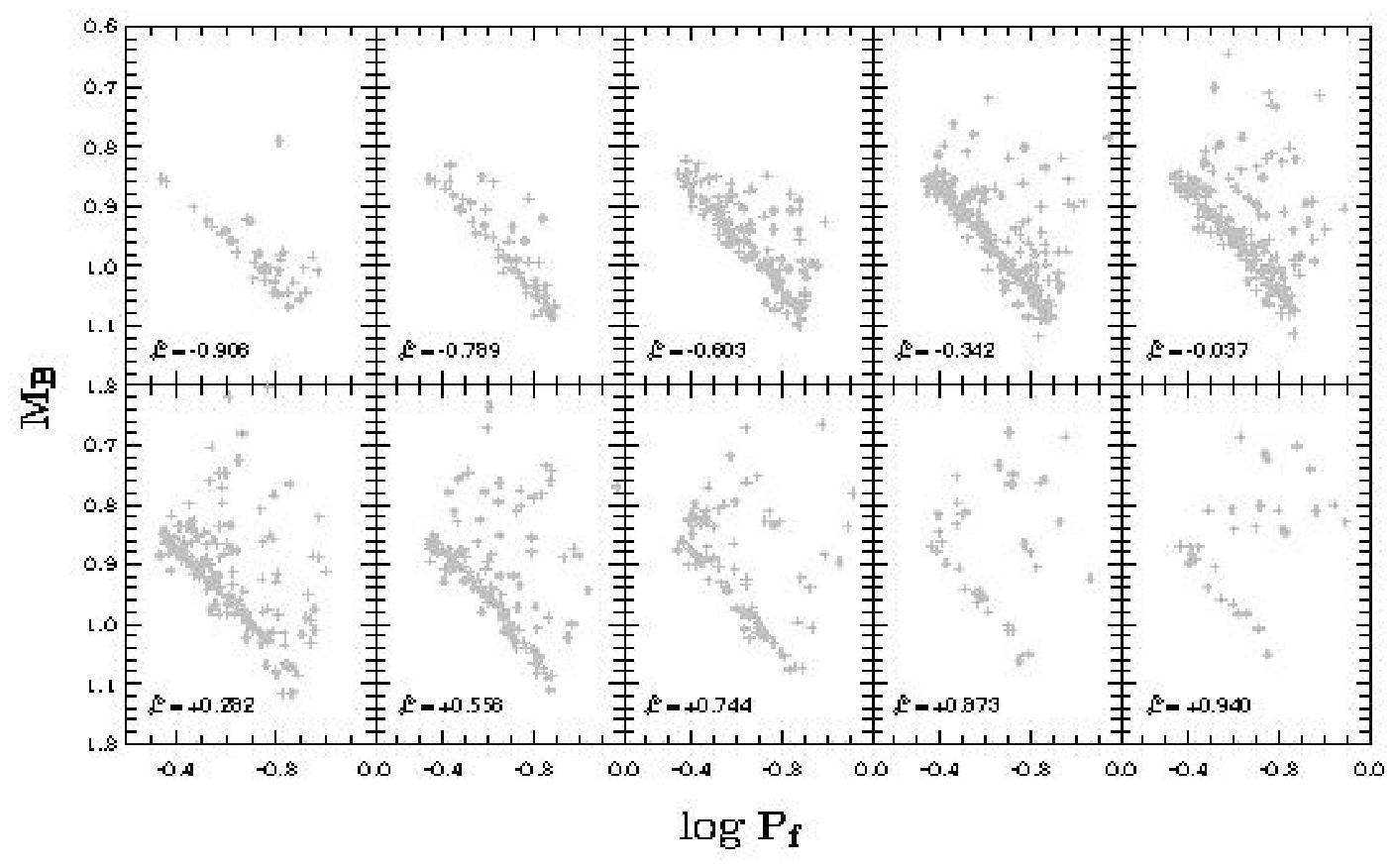}
  \caption{As in Figure~5, but for the $M_B - \log P$ relation.
        }
      \label{Fig6}
\end{sidewaysfigure*}

\begin{sidewaysfigure*}
  \figurenum{7}
  \epsscale{0.875}
  \plotone{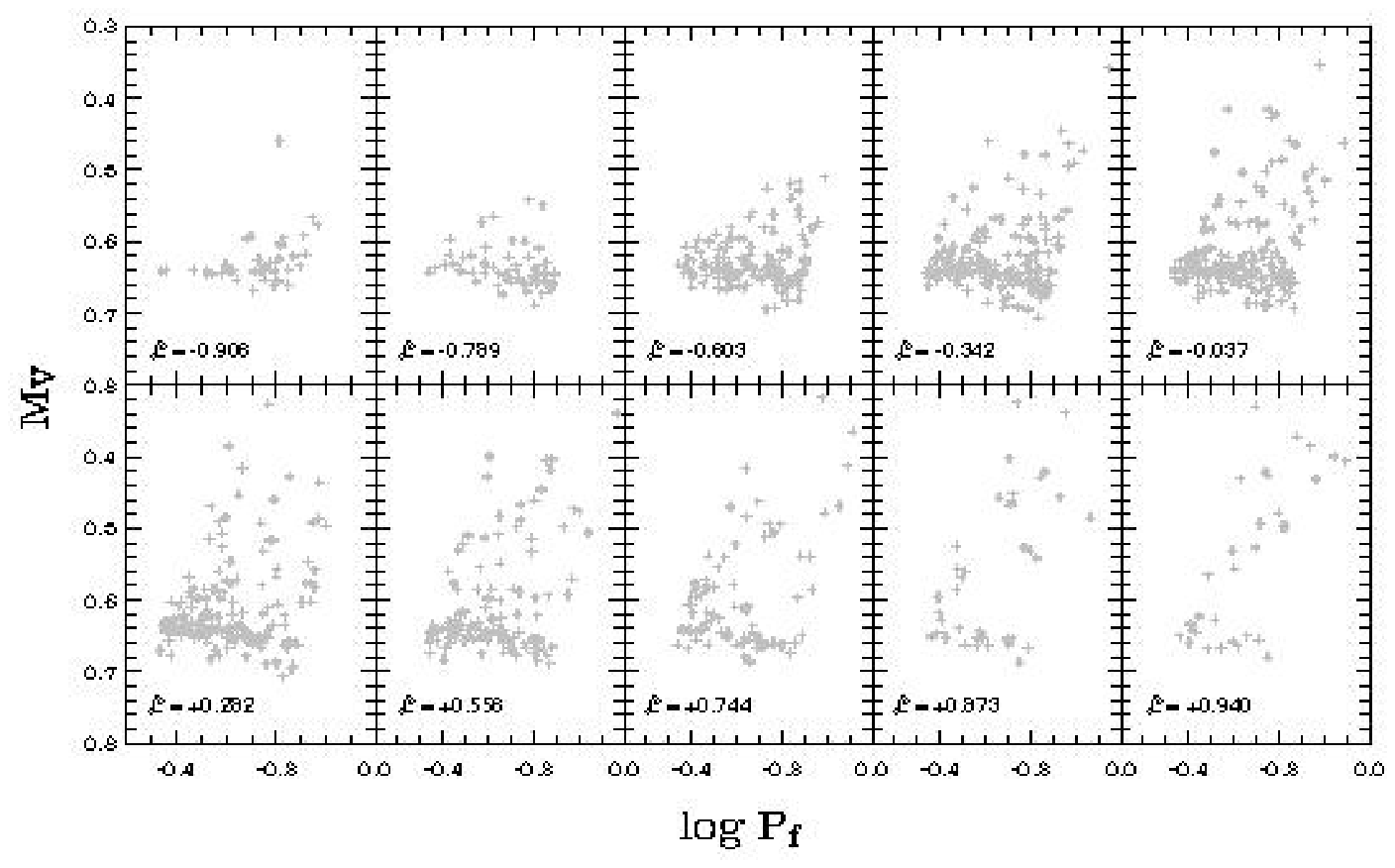}
  \caption{As in Figure~5, but for the $M_V - \log P$ relation.
        }
      \label{Fig7}
\end{sidewaysfigure*}

\begin{sidewaysfigure*}
  \figurenum{8}
  \epsscale{0.875}
  \plotone{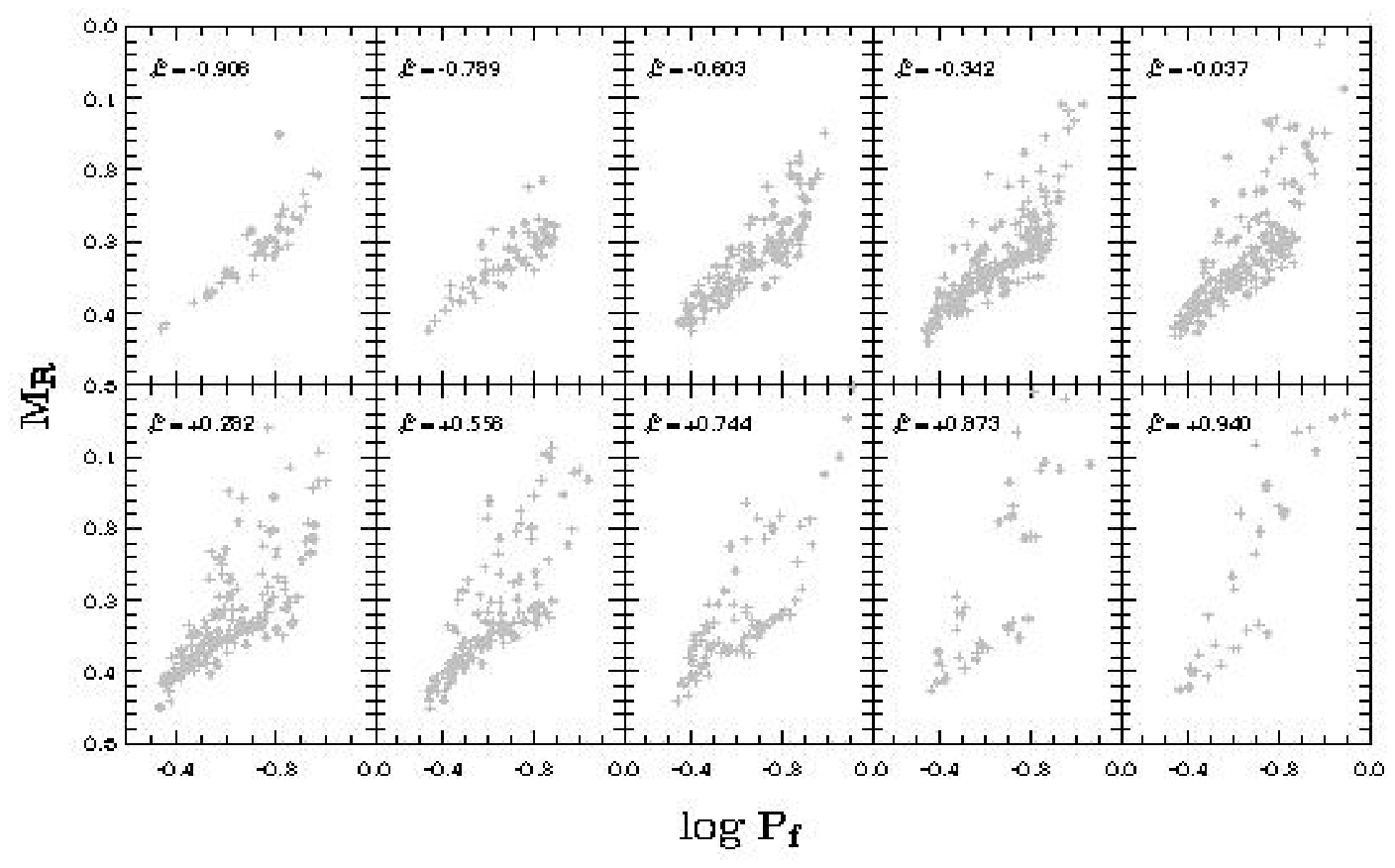}
  \caption{As in Figure~5, but for the $M_R - \log P$ relation.
        }
      \label{Fig8}
\end{sidewaysfigure*}

\begin{sidewaysfigure*}
  \figurenum{9}
  \epsscale{0.875}
  \plotone{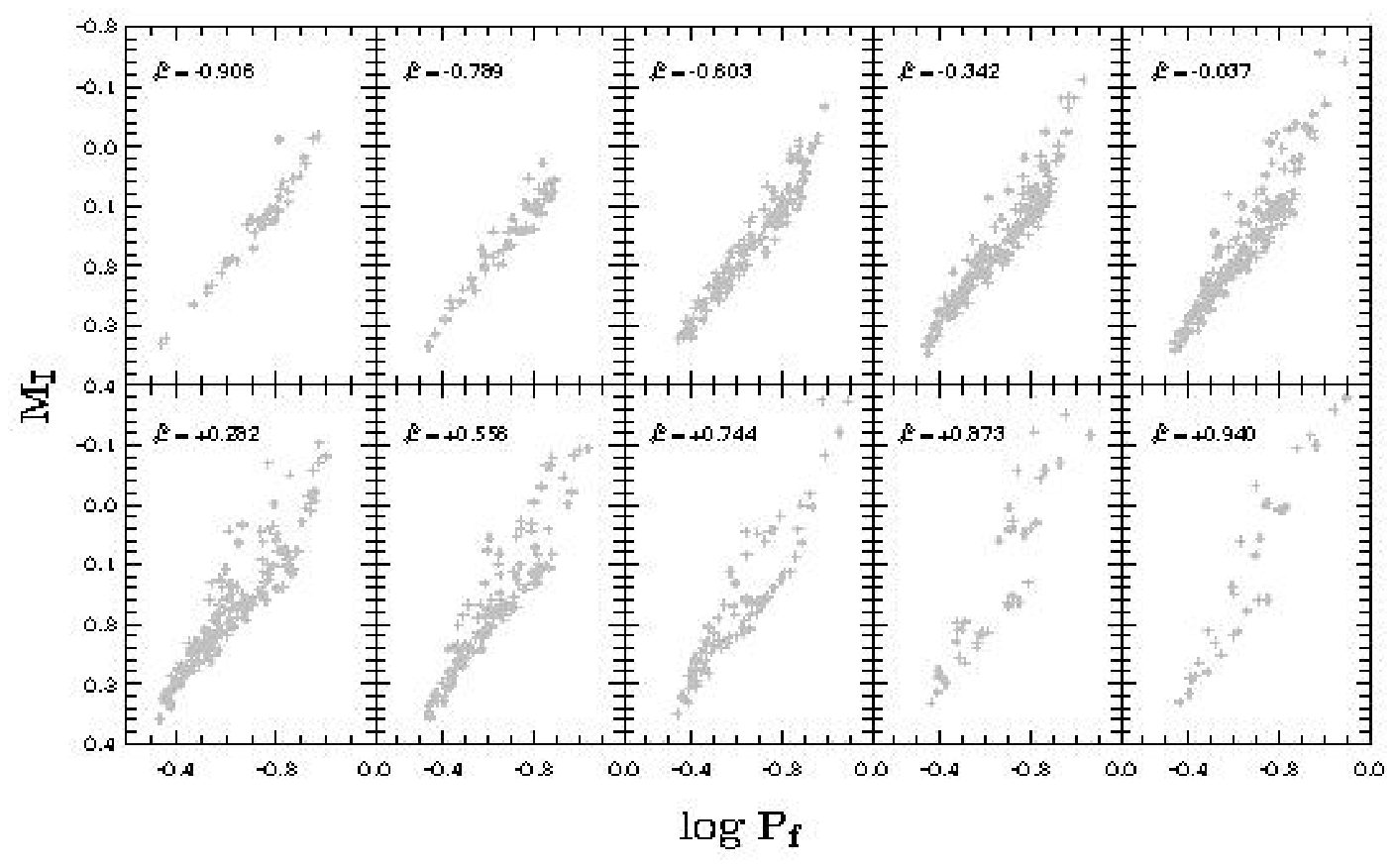}
  \caption{As in Figure~5, but for the $M_I - \log P$ relation.
        }
      \label{Fig9}
\end{sidewaysfigure*}

\begin{sidewaysfigure*}
  \figurenum{10}
  \epsscale{0.875}
  \plotone{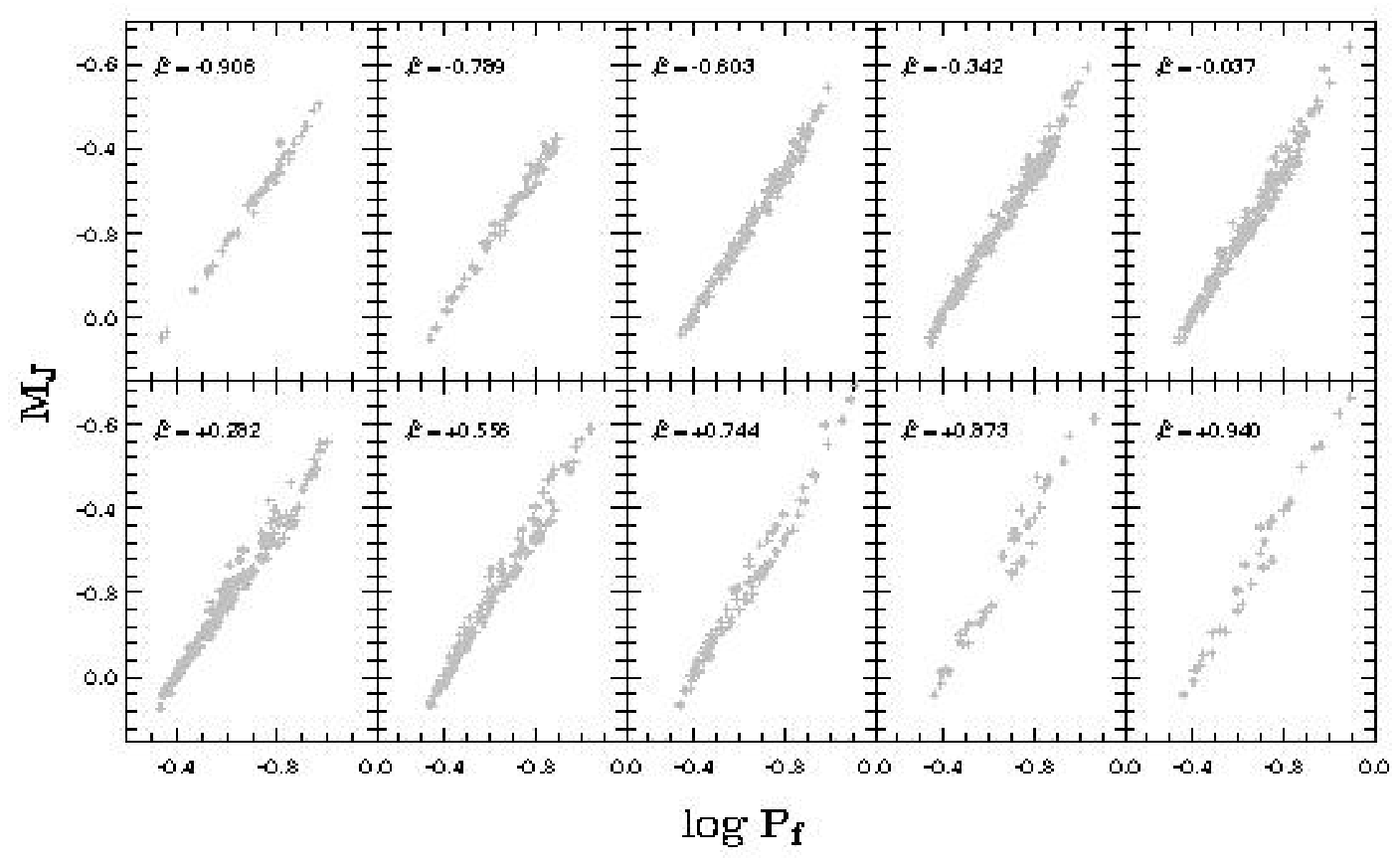}
  \caption{As in Figure~5, but for the $M_J - \log P$ relation.
        }
      \label{Fig10}
\end{sidewaysfigure*}

\begin{sidewaysfigure*}
  \figurenum{11}
  \epsscale{0.875}
  \plotone{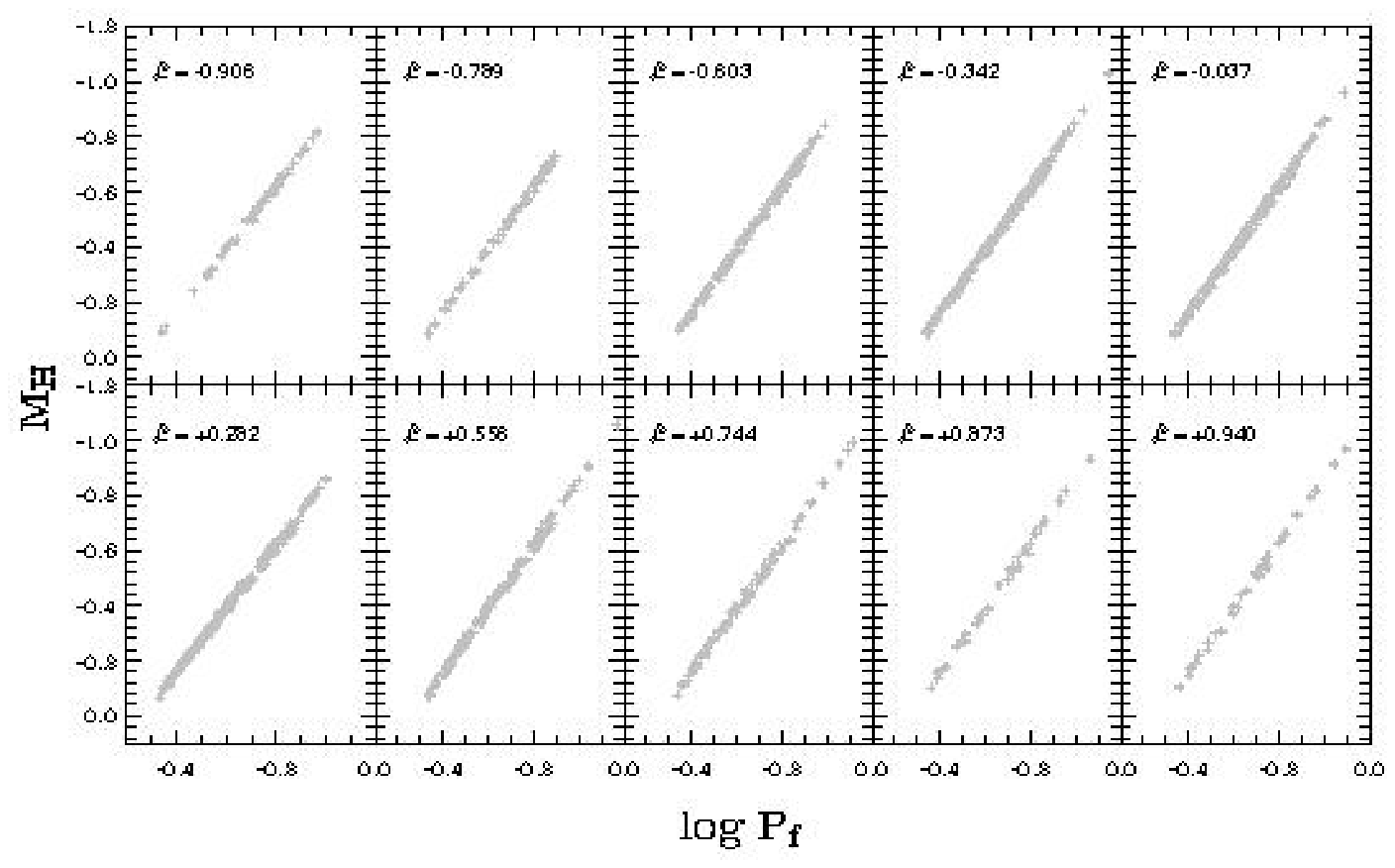}
  \caption{As in Figure~5, but for the $M_H - \log P$ relation.
        }
      \label{Fig11}
\end{sidewaysfigure*}

\begin{sidewaysfigure*}
  \figurenum{12}
  \epsscale{0.875}
  \plotone{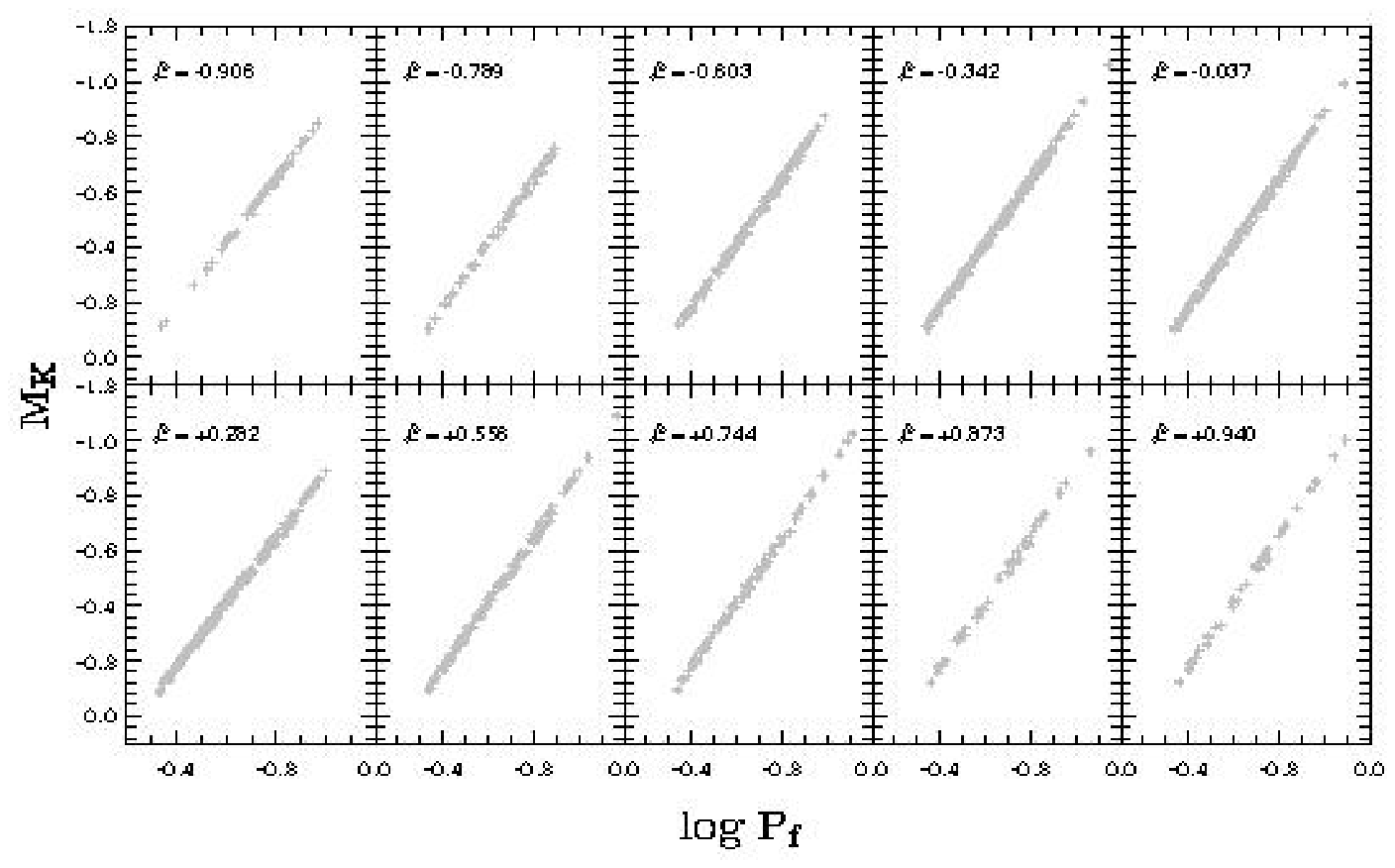}
  \caption{As in Figure~5, but for the $M_K - \log P$ relation.
        }
      \label{Fig12}
\end{sidewaysfigure*}

In order to study the dependence of the zero point and slope of the RRL PL 
relation with both HB type and metallicity, we have computed, for each 
metallicity, sequences of HB simulations which produce from very blue to 
very red HB types. These simulations are standard, and do not include such 
effects as HB bimodality or the impact of second parameters other than 
mass loss on the red giant branch (RGB) or age. 
For each such simulation, linear relations of the type 
$M_X = a + b \, \log P$, in which $X$ represents any of the $UBVRIJHK$ 
bandpasses, were obtained using the Isobe et al. 
(1990) ``OLS bisector'' technique. It is crucial that, if these relations 
are to be compared against empirical data to derive distances, precisely 
the same recipe be employed in the analysis of these data as well, 
particularly in cases in which the correlation coefficient is not very 
close to 1. The final result for each HB morphology actually represents 
the average $a$, $b$ values over 100 HB simulations with 500 stars each.

\begin{deluxetable}{cccccc}  
\tabletypesize{\footnotesize}
\tablecaption{RRL PL Relation in $U$: Coefficients of the Fits} 
\tablewidth{0pt}
\tablehead{
\colhead{$\LeeZinn$} & \colhead{$\sigma({\LeeZinn})$} & \colhead{$a$} & 
\colhead{$\sigma(a)$} & \colhead{$b$} & \colhead{$\sigma(b)$}}
\startdata
\cutinhead{$Z = 0.0005$}
$0.934	$ & $0.013	$ & $0.497	$ & $0.042	$ & $-0.863$ & $	0.129$ \\
$0.877	$ & $0.018	$ & $0.519	$ & $0.034	$ & $-0.849$ & $	0.110$ \\
$0.776	$ & $0.022	$ & $0.547	$ & $0.024	$ & $-0.827$ & $	0.079$ \\
$0.627	$ & $0.027	$ & $0.552	$ & $0.055	$ & $-0.860$ & $	0.192$ \\
$0.414	$ & $0.028	$ & $0.799	$ & $0.267	$ & $-0.037$ & $	0.949$ \\
$0.167	$ & $0.029	$ & $1.045	$ & $0.014	$ & $0.825	$ & $0.058$ \\
$-0.102$ & $	0.031$ & $	1.013$ & $	0.010$ & $	0.706$ & $	0.048$ \\
$-0.358$ & $	0.031$ & $	0.993$ & $	0.010$ & $	0.630$ & $	0.046$ \\
$-0.590$ & $	0.025$ & $	0.983$ & $	0.009$ & $	0.595$ & $	0.041$ \\
$-0.765$ & $	0.021$ & $	0.980$ & $	0.010$ & $	0.586$ & $	0.050$ \\
$-0.883$ & $	0.014$ & $	0.977$ & $	0.013$ & $	0.587$ & $	0.068$ \\
$-0.950$ & $	0.010$ & $	0.979$ & $	0.038$ & $	0.605$ & $	0.200$ \\
\cutinhead{$Z = 0.0010$}
$0.940	$ & $0.011	$ & $0.537	$ & $0.070	$ & $-1.027$ & $	0.191$ \\
$0.873	$ & $0.018	$ & $0.580	$ & $0.070	$ & $-0.952$ & $	0.217$ \\
$0.744	$ & $0.025	$ & $0.609	$ & $0.083	$ & $-0.923$ & $	0.275$ \\
$0.556	$ & $0.034	$ & $0.844	$ & $0.277	$ & $-0.151$ & $	0.962$ \\
$0.282	$ & $0.033	$ & $1.134	$ & $0.109	$ & $0.833	$ & $0.376$ \\
$-0.037$ & $	0.035$ & $	1.136$ & $	0.012$ & $	0.818$ & $	0.055$ \\
$-0.342$ & $	0.036$ & $	1.117$ & $	0.011$ & $	0.741$ & $	0.047$ \\
$-0.603$ & $	0.025$ & $	1.107$ & $	0.012$ & $	0.698$ & $	0.055$ \\
$-0.789$ & $	0.022$ & $	1.098$ & $	0.017$ & $	0.659$ & $	0.081$ \\
$-0.906$ & $	0.015$ & $	1.094$ & $	0.017$ & $	0.646$ & $	0.078$ \\
$-0.963$ & $	0.008$ & $	1.104$ & $	0.034$ & $	0.701$ & $	0.169$ \\
\cutinhead{$Z = 0.0020$}
$0.965	$ & $0.009	$ & $0.675	$ & $0.224	$ & $-0.890$ & $	0.696$ \\
$0.910	$ & $0.015	$ & $0.708	$ & $0.207	$ & $-0.826$ & $	0.643$ \\
$0.794	$ & $0.025	$ & $0.853	$ & $0.286	$ & $-0.405$ & $	0.920$ \\
$0.594	$ & $0.029	$ & $1.155	$ & $0.250	$ & $0.541	$ & $0.816$ \\
$0.307	$ & $0.036	$ & $1.255	$ & $0.117	$ & $0.847	$ & $0.383$ \\
$-0.023$ & $	0.034$ & $	1.268$ & $	0.015$ & $	0.865$ & $	0.066$ \\
$-0.356$ & $	0.035$ & $	1.258$ & $	0.017$ & $	0.820$ & $	0.069$ \\
$-0.630$ & $	0.032$ & $	1.248$ & $	0.017$ & $	0.771$ & $	0.069$ \\
$-0.822$ & $	0.021$ & $	1.247$ & $	0.021$ & $	0.768$ & $	0.089$ \\
$-0.928$ & $	0.012$ & $	1.242$ & $	0.030$ & $	0.739$ & $	0.127$ \\
$-0.974$ & $	0.008$ & $	1.236$ & $	0.077$ & $	0.709$ & $	0.326$ \\
\cutinhead{$Z = 0.0060$}
$0.922	$ & $0.015	$ & $1.032	$ & $0.369	$ & $-0.439$ & $	0.985$ \\
$0.810	$ & $0.021	$ & $1.243	$ & $0.383	$ & $0.128	$ & $1.034$ \\
$0.601	$ & $0.034	$ & $1.445	$ & $0.285	$ & $0.648	$ & $0.792$ \\
$0.298	$ & $0.039	$ & $1.535	$ & $0.186	$ & $0.871	$ & $0.520$ \\
$-0.070$ & $	0.039$ & $	1.574$ & $	0.077$ & $	0.959$ & $	0.210$ \\
$-0.420$ & $	0.036$ & $	1.582$ & $	0.024$ & $	0.958$ & $	0.075$ \\
$-0.693$ & $	0.025$ & $	1.586$ & $	0.032$ & $	0.954$ & $	0.095$ \\
$-0.868$ & $	0.018$ & $	1.571$ & $ 0.079	$ & $0.887	$ & $ 0.229$ \\
$-0.951$ & $	0.012$ & $	1.564$ & $	0.136$ & $	0.866$ & $	0.398$ \\
\hline
\enddata
\end{deluxetable}

\section{Genesis of the RRL PL Relation} 
In Figure~1, we show an HB simulation computed for a metallicity $Z = 0.002$ 
and an intermediate HB morphology, indicated by a value of the Lee-Zinn type 
$\LeeZinn \equiv (B-R)/(B+V+R) = -0.05$ (where $B$, $V$, and $R$ are the 
numbers of blue, 
variable, and red HB stars, respectively). Even using only the more usual 
$BVI$ bandpasses of the Johnson-Cousins system, the change in the detailed 
morphology of the HB with the passband adopted is obvious. In the middle 
upper panel, one can see the traditional display of a ``horizontal'' HB, 
as obtained in the $M_V$, $\bv$ plane. As a consequence, one can see, in 
the middle lower panel, that no PL relation results using this bandpass. On 
the other hand, the upper left panel shows the same simulation in the $M_B$, 
$\bv$ plane. One clearly sees now that the HB is not anymore ``horizontal.''
This has a clear impact upon the resulting PL relation (lower left panel): 
now one does see an indication of a correlation between period and $M_B$, 
though with a large scatter. The reason for this scatter is that the 
effects of luminosity and temperature variations upon the expected periods 
are almost orthogonal in this plane. Now one can also see, in the upper 
right panel, that the HB is also not quite horizontal in the $M_I$, $\bv$ 
plane---only that now, in comparison with the $M_B$, $\bv$ plane, the 
stars that look brighter are also the ones that are cooler. Since a 
decrease in temperature, as well as an increase in brightness, both lead 
to longer periods, one expects the effects of brightness and temperature 
upon the periods to be more nearly parallel when using $I$. This is indeed 
what happens, as can be seen in the bottom right panel. We now find a quite 
reasonable PL relation, with much less scatter than was the case in $B$.

\begin{deluxetable}{cccccc}  
\tabletypesize{\footnotesize}
\tablecaption{RRL PL Relation in $B$: Coefficients of the Fits} 
\tablewidth{0pt}
\tablehead{
\colhead{$\LeeZinn$} & \colhead{$\sigma({\LeeZinn})$} & \colhead{$a$} & 
\colhead{$\sigma(a)$} & \colhead{$b$} & \colhead{$\sigma(b)$}}
\startdata
\cutinhead{$Z = 0.0005$}
$0.934	$ & $0.013	$ & $0.553	$ & $0.124	$ & $-0.763$ & $	0.524$ \\
$0.877	$ & $0.018	$ & $0.555	$ & $0.110	$ & $-0.838$ & $	0.431$ \\
$0.776	$ & $0.022	$ & $0.739	$ & $0.257	$ & $-0.263$ & $	0.940$ \\
$0.627	$ & $0.027	$ & $1.093	$ & $0.058	$ & $0.933	$ & $0.199$ \\
$0.414	$ & $0.028	$ & $1.104	$ & $0.013	$ & $0.899	$ & $0.044$ \\
$0.167	$ & $0.029	$ & $1.107	$ & $0.012	$ & $0.873	$ & $0.040$ \\
$-0.102$ & $	0.031$ & $	1.106$ & $	0.010$ & $	0.851$ & $	0.037$ \\
$-0.358$ & $	0.031$ & $	1.105$ & $	0.010$ & $	0.836$ & $	0.035$ \\
$-0.590$ & $	0.025$ & $	1.105$ & $	0.010$ & $	0.831$ & $	0.036$ \\
$-0.765$ & $	0.021$ & $	1.107$ & $	0.011$ & $	0.839$ & $	0.047$ \\
$-0.883$ & $	0.014$ & $	1.105$ & $	0.013$ & $	0.838$ & $	0.060$ \\
$-0.950$ & $	0.010$ & $	1.108$ & $	0.020$ & $	0.853$ & $	0.100$ \\
\cutinhead{$Z = 0.0010$}
$0.940	$ & $0.011	$ & $0.568	$ & $0.123	$ & $-0.966$ & $	0.437$ \\
$0.873	$ & $0.018	$ & $0.651	$ & $0.196	$ & $-0.768$ & $	0.665$ \\
$0.744	$ & $0.025	$ & $0.994	$ & $0.271	$ & $0.362	$ & $0.944$ \\
$0.556	$ & $0.034	$ & $1.195	$ & $0.016	$ & $1.001	$ & $0.048$ \\
$0.282	$ & $0.033	$ & $1.202	$ & $0.013	$ & $0.973	$ & $0.046$ \\
$-0.037$ & $	0.035$ & $	1.206$ & $	0.015$ & $	0.955$ & $	0.051$ \\
$-0.342$ & $	0.036$ & $	1.205$ & $	0.013$ & $	0.935$ & $	0.044$ \\
$-0.603$ & $	0.025$ & $	1.205$ & $	0.013$ & $	0.923$ & $	0.048$ \\
$-0.789$ & $	0.022$ & $	1.203$ & $	0.015$ & $	0.906$ & $	0.067$ \\
$-0.906$ & $	0.015$ & $	1.200$ & $	0.019$ & $	0.895$ & $	0.076$ \\
$-0.963$ & $	0.008$ & $	1.207$ & $	0.036$ & $	0.930$ & $	0.164$ \\
\cutinhead{$Z = 0.0020$}
$0.965	$ & $0.009	$ & $0.741	$ & $0.300	$ & $-0.659$ & $	0.974$ \\
$0.910	$ & $0.015	$ & $0.822	$ & $0.305	$ & $-0.442$ & $	0.972$ \\
$0.794	$ & $0.025	$ & $1.116	$ & $0.294	$ & $0.466	$ & $0.939$ \\
$0.594	$ & $0.029	$ & $1.302	$ & $0.068	$ & $1.022	$ & $0.214$ \\
$0.307	$ & $0.036	$ & $1.315	$ & $0.017	$ & $1.024	$ & $0.059$ \\
$-0.023$ & $	0.034$ & $	1.320$ & $	0.017$ & $	1.011$ & $	0.063$ \\
$-0.356$ & $	0.035$ & $	1.319$ & $	0.018$ & $	0.990$ & $	0.066$ \\
$-0.630$ & $	0.032$ & $	1.320$ & $	0.019$ & $	0.969$ & $	0.066$ \\
$-0.822$ & $	0.021$ & $	1.320$ & $	0.021$ & $	0.969$ & $	0.080$ \\
$-0.928$ & $	0.012$ & $	1.321$ & $	0.033$ & $	0.955$ & $	0.125$ \\
$-0.974$ & $	0.008$ & $	1.315$ & $	0.084$ & $	0.923$ & $	0.346$ \\
\cutinhead{$Z = 0.0060$}
$0.922	$ & $0.015	$ & $1.072	$ & $0.413	$ & $-0.205$ & $	1.110$ \\
$0.810	$ & $0.021	$ & $1.288	$ & $0.377	$ & $0.370	$ & $1.020$ \\
$0.601	$ & $0.034	$ & $1.495	$ & $0.212	$ & $0.907	$ & $0.589$ \\
$0.298	$ & $0.039	$ & $1.563	$ & $0.079	$ & $1.067	$ & $0.222$ \\
$-0.070$ & $	0.039$ & $	1.574$ & $	0.027$ & $	1.068$ & $	0.069$ \\
$-0.420$ & $	0.036$ & $	1.580$ & $	0.028$ & $	1.060$ & $	0.078$ \\
$-0.693$ & $	0.025$ & $	1.588$ & $	0.035$ & $	1.065$ & $	0.099$ \\
$-0.868$ & $	0.018$ & $	1.573$ & $	0.108$ & $	0.992$ & $	0.317$ \\
$-0.951$ & $	0.012$ & $	1.571$ & $	0.145$ & $	0.987$ & $	0.420$ \\
\hline
\enddata
\end{deluxetable}

The same concepts explain the behavior of the RRL PL relation in the 
other passbands of the Johnson-Cousins-Glass system, which becomes tighter 
both towards the near-ultraviolet and towards the near-infrared, as compared 
to the visual. In Figure~2, we show the PL relations in all of the $UBVR$ 
(upper panels) and $IJHK$ (lower panels) bandpasses, for a synthetic HB 
with a morphology similar to that shown in Figure~1, but computed for a 
metallicity $Z = 0.001$ (the results are qualitatively similar for all 
metallicities). As one can see, as one moves redward from $V$, where the 
HB is effectively horizontal at the RRL level, an increasingly tighter 
PL relation develops. Conversely, as one moves from $V$ towards the 
ultraviolet, the expectation is also for the PL relation to become 
increasingly tighter---which is confirmed by the plot for $B$. In the 
case of broadband $U$, as can be seen, the expected tendency is not 
fully confirmed, an effect which we attribute to the complicating 
impact of the Balmer jump upon the predicted bolometric 
corrections in the region of interest.\footnote{The Balmer jump 
occurs at around $\lambda \approx 3700 \, {\rm \AA}$, marking the asymptotic 
end of the Balmer line series---and thus a discontinuity in the 
radiative opacity. The broadband $U$ filter extends well redward of 
$4000\, {\rm \AA}$, and is thus strongly affected by the detailed physics 
controlling the size of the Balmer jump. In the case of Str\"omgren $u$, 
on the other hand, the transmission efficiency is practically zero 
already at $\lambda = 3800 \, {\rm \AA}$, thus showing that it is not severely 
affected by the size of the Balmer jump.} 
An investigation of the RRL PL 
relation in Str\"omgren $u$ (e.g., Clem et al. 2004), which is much less 
affected by the Balmer discontinuity (and might accordingly produce 
a tighter PL relation than in broadband $U$), as well as of the UV domain, 
should thus prove of interest, but has not been attempted in the present 
work.

\begin{deluxetable}{cccccc}  
\tabletypesize{\footnotesize}
\tablecaption{RRL PL Relation in $V$: Coefficients of the Fits} 
\tablewidth{0pt}
\tablehead{
\colhead{$\LeeZinn$} & \colhead{$\sigma({\LeeZinn})$} & \colhead{$a$} & 
\colhead{$\sigma(a)$} & \colhead{$b$} & \colhead{$\sigma(b)$}}
\startdata
\cutinhead{$Z = 0.0005$}
$0.934	$ & $0.013	$ & $0.203	$ & $0.038	$ & $-0.973$ & $	0.126$ \\
$0.877	$ & $0.018	$ & $0.231	$ & $0.029	$ & $-0.947$ & $	0.096$ \\
$0.776	$ & $0.022	$ & $0.274	$ & $0.021	$ & $-0.871$ & $	0.065$ \\
$0.627	$ & $0.027	$ & $0.305	$ & $0.017	$ & $-0.815$ & $	0.049$ \\
$0.414	$ & $0.028	$ & $0.338	$ & $0.013	$ & $-0.749$ & $	0.041$ \\
$0.167	$ & $0.029	$ & $0.360	$ & $0.013	$ & $-0.723$ & $	0.045$ \\
$-0.102$ & $	0.031$ & $	0.374$ & $	0.014$ & $	-0.714$ & $	0.053$ \\
$-0.358$ & $	0.031$ & $	0.385$ & $	0.017$ & $	-0.723$ & $	0.069$ \\
$-0.590$ & $	0.025$ & $	0.415$ & $	0.096$ & $	-0.639$ & $	0.409$ \\
$-0.765$ & $	0.021$ & $	0.448$ & $	0.134$ & $	-0.530$ & $	0.600$ \\
$-0.883$ & $	0.014$ & $	0.510$ & $	0.158$ & $	-0.271$ & $	0.745$ \\
$-0.950$ & $	0.010$ & $	0.520$ & $	0.137$ & $	-0.249$ & $	0.689$ \\
\cutinhead{$Z = 0.0010$}
$0.940	$ & $0.011	$ & $0.215	$ & $0.065	$ & $-1.162$ & $	0.188$ \\
$0.873	$ & $0.018	$ & $0.265	$ & $0.036	$ & $-1.060$ & $	0.103$ \\
$0.744	$ & $0.025	$ & $0.312	$ & $0.028	$ & $-0.967$ & $	0.088$ \\
$0.556	$ & $0.034	$ & $0.351	$ & $0.019	$ & $-0.879$ & $	0.057$ \\
$0.282	$ & $0.033	$ & $0.383	$ & $0.016	$ & $-0.815$ & $	0.044$ \\
$-0.037$ & $	0.035$ & $	0.399$ & $	0.016$ & $	-0.801$ & $	0.057$ \\
$-0.342$ & $	0.036$ & $	0.428$ & $	0.096$ & $	-0.741$ & $	0.359$ \\
$-0.603$ & $	0.025$ & $	0.468$ & $	0.147$ & $	-0.627$ & $	0.567$ \\
$-0.789$ & $	0.022$ & $	0.596$ & $	0.205$ & $	-0.146$ & $	0.824$ \\
$-0.906$ & $	0.015$ & $	0.625$ & $	0.183$ & $	-0.037$ & $	0.762$ \\
$-0.963$ & $	0.008$ & $	0.664$ & $	0.152$ & $	0.126$ & $	0.678$ \\
\cutinhead{$Z = 0.0020$}
$0.965	$ & $0.009	$ & $0.276	$ & $0.142	$ & $-1.195$ & $	0.415$ \\
$0.910	$ & $0.015	$ & $0.310	$ & $0.063	$ & $-1.118$ & $	0.172$ \\
$0.794	$ & $0.025	$ & $0.361	$ & $0.041	$ & $-1.007$ & $	0.107$ \\
$0.594	$ & $0.029	$ & $0.401	$ & $0.024	$ & $-0.919$ & $	0.068$ \\
$0.307	$ & $0.036	$ & $0.429	$ & $0.023	$ & $-0.865$ & $	0.061$ \\
$-0.023$ & $	0.034$ & $	0.464$ & $	0.097$ & $	-0.784$ & $	0.314$ \\
$-0.356$ & $	0.035$ & $	0.477$ & $	0.092$ & $	-0.771$ & $	0.312$ \\
$-0.630$ & $	0.032$ & $	0.533$ & $	0.171$ & $	-0.610$ & $	0.590$ \\
$-0.822$ & $	0.021$ & $	0.630$ & $	0.224$ & $	-0.293$ & $	0.796$ \\
$-0.928$ & $	0.012$ & $	0.733$ & $	0.196$ & $	0.072$ & $	0.715$ \\
$-0.974$ & $	0.008$ & $	0.737$ & $	0.170$ & $	0.073$ & $	0.654$ \\
\cutinhead{$Z = 0.0060$}
$0.922	$ & $0.015	$ & $0.428	$ & $0.177	$ & $-1.103$ & $	0.445$ \\
$0.810	$ & $0.021	$ & $0.454	$ & $0.065	$ & $-1.061$ & $	0.152$ \\
$0.601	$ & $0.034	$ & $0.487	$ & $0.046	$ & $-1.001$ & $	0.113$ \\
$0.298	$ & $0.039	$ & $0.513	$ & $0.033	$ & $-0.957$ & $	0.084$ \\
$-0.070$ & $	0.039$ & $	0.551$ & $	0.074$ & $	-0.880$ & $	0.201$ \\
$-0.420$ & $	0.036$ & $	0.584$ & $	0.117$ & $	-0.815$ & $	0.325$ \\
$-0.693$ & $	0.025$ & $	0.642$ & $	0.201$ & $	-0.677$ & $	0.561$ \\
$-0.868$ & $	0.018$ & $	0.789$ & $	0.273$ & $	-0.280$ & $	0.781$ \\
$-0.951$ & $	0.012$ & $	0.840$ & $	0.257$ & $	-0.142$ & $	0.748$ \\ 
\hline
\enddata
\end{deluxetable}

\section{The RRL PL Relation Calibrated} 
In Figure~3, we show the slope (left panels) and zero point (right panels) 
of the theoretically-calibrated RRL PL relation, in $UBVR$ (from top to 
bottom) and for four different metallicities (as indicated by different 
symbols and shades of gray; see the lower right panel). Each datapoint 
corresponds to the average over 100 simulations with 500 stars in each. 
The ``error bars'' correspond to the standard deviation of the mean over 
these 100 simulations. Figure~4 is analogous to Figure~3, but shows 
instead our results for the $IJHK$ passbands (from top to bottom). 
It should be noted that, for all 
bandpasses, the coefficients of the PL relations are much more subject 
to statistical fluctuations at the extremes in HB type (both very red 
and very blue), due to the smaller numbers of RRL variables for these 
HB types. In terms of Figures~3 and 4, this is indicated by an increase 
in the size of the ``error bars'' at both the blue and red ends of the 
relations.

\begin{deluxetable}{cccccc}  
\tabletypesize{\footnotesize}
\tablecaption{RRL PL Relation in $R$: Coefficients of the Fits} 
\tablewidth{0pt}
\tablehead{
\colhead{$\LeeZinn$} & \colhead{$\sigma({\LeeZinn})$} & \colhead{$a$} & 
\colhead{$\sigma(a)$} & \colhead{$b$} & \colhead{$\sigma(b)$}}
\startdata
\cutinhead{$Z = 0.0005$}
$0.934	$ & $0.013	$ & $-0.132$ & $	0.031$ & $	-1.195$ & $	0.108$ \\
$0.877	$ & $0.018	$ & $-0.106$ & $	0.021$ & $	-1.163$ & $	0.068$ \\
$0.776	$ & $0.022	$ & $-0.065$ & $	0.017$ & $	-1.081$ & $	0.049$ \\
$0.627	$ & $0.027	$ & $-0.028$ & $	0.014$ & $	-0.996$ & $	0.040$ \\
$0.414	$ & $0.028	$ & $0.013	$ & $0.011	$ & $-0.891$ & $	0.033$ \\
$0.167	$ & $0.029	$ & $0.047	$ & $0.011	$ & $-0.801$ & $	0.033$ \\
$-0.102$ & $	0.031$ & $	0.070$ & $	0.009$ & $	-0.738$ & $	0.031$ \\
$-0.358$ & $	0.031$ & $	0.087$ & $	0.008$ & $	-0.689$ & $	0.029$ \\
$-0.590$ & $	0.025$ & $	0.098$ & $	0.008$ & $	-0.656$ & $	0.032$ \\
$-0.765$ & $	0.021$ & $	0.107$ & $	0.010$ & $	-0.630$ & $	0.039$ \\
$-0.883$ & $	0.014$ & $	0.110$ & $	0.012$ & $	-0.619$ & $	0.055$ \\
$-0.950$ & $	0.010$ & $	0.110$ & $	0.020$ & $	-0.634$ & $	0.102$ \\
\cutinhead{$Z = 0.0010$}
$0.940	$ & $0.011	$ & $-0.119$ & $	0.051$ & $	-1.354$ & $	0.148$ \\
$0.873	$ & $0.018	$ & $-0.070$ & $	0.032$ & $	-1.248$ & $	0.089$ \\
$0.744	$ & $0.025	$ & $-0.021$ & $	0.026$ & $	-1.134$ & $	0.078$ \\
$0.556	$ & $0.034	$ & $0.023	$ & $0.018	$ & $-1.017$ & $	0.054$ \\
$0.282	$ & $0.033	$ & $0.065	$ & $0.017	$ & $-0.906$ & $	0.051$ \\
$-0.037$ & $	0.035$ & $	0.098$ & $	0.015$ & $	-0.815$ & $	0.047$ \\
$-0.342$ & $	0.036$ & $	0.125$ & $	0.014$ & $	-0.740$ & $	0.045$ \\
$-0.603$ & $	0.025$ & $	0.143$ & $	0.014$ & $	-0.687$ & $	0.049$ \\
$-0.789$ & $	0.022$ & $	0.159$ & $	0.016$ & $	-0.639$ & $	0.058$ \\
$-0.906$ & $	0.015$ & $	0.167$ & $	0.019$ & $	-0.610$ & $	0.070$ \\
$-0.963$ & $	0.008$ & $	0.167$ & $	0.027$ & $	-0.623$ & $	0.116$ \\
\cutinhead{$Z = 0.0020$}
$0.965	$ & $0.009	$ & $-0.071$ & $	0.092$ & $	-1.401$ & $	0.268$ \\
$0.910	$ & $0.015	$ & $-0.027$ & $	0.059$ & $	-1.290$ & $	0.163$ \\
$0.794	$ & $0.025	$ & $0.027	$ & $0.043	$ & $-1.158$ & $	0.118$ \\
$0.594	$ & $0.029	$ & $0.074	$ & $0.028	$ & $-1.039$ & $	0.082$ \\
$0.307	$ & $0.036	$ & $0.110	$ & $0.029	$ & $-0.948$ & $	0.082$ \\
$-0.023$ & $	0.034$ & $	0.145$ & $	0.024$ & $	-0.854$ & $	0.070$ \\
$-0.356$ & $	0.035$ & $	0.165$ & $	0.020$ & $	-0.803$ & $	0.061$ \\
$-0.630$ & $	0.032$ & $	0.191$ & $	0.022$ & $	-0.732$ & $	0.067$ \\
$-0.822$ & $	0.021$ & $	0.201$ & $	0.029$ & $	-0.708$ & $	0.095$ \\
$-0.928$ & $	0.012$ & $	0.224$ & $	0.036$ & $	-0.640$ & $	0.116$ \\
$-0.974$ & $	0.008$ & $	0.233$ & $	0.049$ & $	-0.619$ & $	0.183$ \\
\cutinhead{$Z = 0.0060$}
$0.922	$ & $0.015	$ & $0.060	$ & $0.103	$ & $-1.332$ & $	0.252$ \\
$0.810	$ & $0.021	$ & $0.111	$ & $0.070	$ & $-1.217$ & $	0.168$ \\
$0.601	$ & $0.034	$ & $0.147	$ & $0.051	$ & $-1.142$ & $	0.127$ \\
$0.298	$ & $0.039	$ & $0.179	$ & $0.041	$ & $-1.076$ & $	0.106$ \\
$-0.070$ & $	0.039$ & $	0.217$ & $	0.041$ & $	-0.991$ & $	0.103$ \\
$-0.420$ & $	0.036$ & $	0.246$ & $	0.039$ & $	-0.928$ & $	0.101$ \\
$-0.693$ & $	0.025$ & $	0.270$ & $	0.048$ & $	-0.876$ & $	0.125$ \\
$-0.868$ & $	0.018$ & $	0.315$ & $	0.055$ & $	-0.767$ & $	0.146$ \\
$-0.951$ & $	0.012$ & $	0.320$ & $	0.080$ & $	-0.763$ & $	0.221$ \\
\hline
\enddata
\end{deluxetable}

The slopes and zero points for the $UBVRIJHK$ calibrations are given 
in Tables~1 through 8, respectively. Appropriate values 
for any given 
HB morphology may be obtained from these tables by direct interpolation, 
or by using suitable interpolation formulae (Catelan 2004b), which we 
now proceed to describe in more detail.

\begin{deluxetable}{cccccc}  
\tabletypesize{\footnotesize}
\tablecaption{RRL PL Relation in $I$: Coefficients of the Fits} 
\tablewidth{0pt}
\tablehead{
\colhead{$\LeeZinn$} & \colhead{$\sigma({\LeeZinn})$} & \colhead{$a$} & 
\colhead{$\sigma(a)$} & \colhead{$b$} & \colhead{$\sigma(b)$}}
\startdata
\cutinhead{$Z = 0.0005$}
$0.934	$ & $0.013	$ & $-0.343$ & $	0.023$ & $	-1.453$ & $	0.082$ \\
$0.877	$ & $0.018	$ & $-0.322$ & $	0.015$ & $	-1.426$ & $	0.046$ \\
$0.776	$ & $0.022	$ & $-0.291$ & $	0.012$ & $	-1.364$ & $	0.036$ \\
$0.627	$ & $0.027	$ & $-0.261$ & $	0.010$ & $	-1.294$ & $	0.030$ \\
$0.414	$ & $0.028	$ & $-0.231$ & $	0.009$ & $	-1.215$ & $	0.027$ \\
$0.167	$ & $0.029	$ & $-0.207$ & $	0.008$ & $	-1.150$ & $	0.026$ \\
$-0.102$ & $	0.031$ & $	-0.192$ & $	0.007$ & $	-1.109$ & $	0.022$ \\
$-0.358$ & $	0.031$ & $	-0.182$ & $	0.006$ & $	-1.080$ & $	0.020$ \\
$-0.590$ & $	0.025$ & $	-0.176$ & $	0.006$ & $	-1.060$ & $	0.023$ \\
$-0.765$ & $	0.021$ & $	-0.171$ & $	0.007$ & $	-1.044$ & $	0.029$ \\
$-0.883$ & $	0.014$ & $	-0.170$ & $	0.008$ & $	-1.038$ & $	0.037$ \\
$-0.950$ & $	0.010$ & $	-0.171$ & $	0.014$ & $	-1.050$ & $	0.070$ \\
\cutinhead{$Z = 0.0010$}
$0.940	$ & $0.011	$ & $-0.318$ & $	0.037$ & $	-1.568$ & $	0.107$ \\
$0.873	$ & $0.018	$ & $-0.279$ & $	0.025$ & $	-1.482$ & $	0.068$ \\
$0.744	$ & $0.025	$ & $-0.239$ & $	0.019$ & $	-1.385$ & $	0.057$ \\
$0.556	$ & $0.034	$ & $-0.204$ & $	0.014$ & $	-1.292$ & $	0.042$ \\
$0.282	$ & $0.033	$ & $-0.172$ & $	0.013$ & $	-1.208$ & $	0.039$ \\
$-0.037$ & $	0.035$ & $	-0.148$ & $	0.012$ & $	-1.140$ & $	0.037$ \\
$-0.342$ & $	0.036$ & $	-0.130$ & $	0.011$ & $	-1.089$ & $	0.034$ \\
$-0.603$ & $	0.025$ & $	-0.119$ & $	0.010$ & $	-1.055$ & $	0.033$ \\
$-0.789$ & $	0.022$ & $	-0.110$ & $	0.011$ & $	-1.028$ & $	0.037$ \\
$-0.906$ & $	0.015$ & $	-0.106$ & $	0.014$ & $	-1.012$ & $	0.051$ \\
$-0.963$ & $	0.008$ & $	-0.105$ & $	0.020$ & $	-1.013$ & $	0.086$ \\
\cutinhead{$Z = 0.0020$}
$0.965	$ & $0.009	$ & $-0.265$ & $	0.074$ & $	-1.592$ & $	0.216$ \\
$0.910	$ & $0.015	$ & $-0.230$ & $	0.047$ & $	-1.503$ & $	0.128$ \\
$0.794	$ & $0.025	$ & $-0.185$ & $	0.035$ & $	-1.390$ & $	0.097$ \\
$0.594	$ & $0.029	$ & $-0.148$ & $	0.023$ & $	-1.295$ & $	0.067$ \\
$0.307	$ & $0.036	$ & $-0.120$ & $	0.022$ & $	-1.225$ & $	0.062$ \\
$-0.023$ & $	0.034$ & $	-0.094$ & $	0.018$ & $	-1.154$ & $	0.052$ \\
$-0.356$ & $	0.035$ & $	-0.080$ & $	0.014$ & $	-1.119$ & $	0.044$ \\
$-0.630$ & $	0.032$ & $	-0.062$ & $	0.016$ & $	-1.070$ & $	0.048$ \\
$-0.822$ & $	0.021$ & $	-0.056$ & $	0.020$ & $	-1.054$ & $	0.066$ \\
$-0.928$ & $	0.012$ & $	-0.041$ & $	0.025$ & $	-1.011$ & $	0.080$ \\
$-0.974$ & $	0.008$ & $	-0.036$ & $	0.037$ & $	-0.999$ & $	0.137$ \\
\cutinhead{$Z = 0.0060$}
$0.922	$ & $0.015	$ & $-0.142$ & $	0.088$ & $	-1.523$ & $	0.214$ \\
$0.810	$ & $0.021	$ & $-0.098$ & $	0.057$ & $	-1.423$ & $	0.139$ \\
$0.601	$ & $0.034	$ & $-0.068$ & $	0.042$ & $	-1.358$ & $	0.104$ \\
$0.298	$ & $0.039	$ & $-0.042$ & $	0.034$ & $	-1.302$ & $	0.086$ \\
$-0.070$ & $	0.039$ & $	-0.013$ & $	0.033$ & $	-1.237$ & $	0.083$ \\
$-0.420$ & $	0.036$ & $	0.010$ & $	0.031$ & $	-1.188$ & $	0.080$ \\
$-0.693$ & $	0.025$ & $	0.030$ & $	0.037$ & $	-1.145$ & $	0.098$ \\
$-0.868$ & $	0.018$ & $	0.062$ & $	0.042$ & $	-1.068$ & $	0.112$ \\
$-0.951$ & $	0.012$ & $	0.066$ & $	0.061$ & $	-1.061$ & $	0.168$ \\
\hline
\enddata
\end{deluxetable}

\subsection{Analytical Fits} 
As the plots in Figures~3 and 4 show, all bands show some dependence on 
both metallicity and HB type, though some of the effects clearly become 
less pronounced as one goes towards the near-infrared. Analysis of 
the data for each metallicity shows that, except for the $U$ and $B$ 
cases, the coefficients of all PL relations (at a fixed metallicity) 
can be well described by third-order polynomials, as follows:

\begin{deluxetable}{cccccc}  
\tabletypesize{\footnotesize}
\tablecaption{RRL PL Relation in $J$: Coefficients of the Fits} 
\tablewidth{0pt}
\tablehead{
\colhead{$\LeeZinn$} & \colhead{$\sigma({\LeeZinn})$} & \colhead{$a$} & 
\colhead{$\sigma(a)$} & \colhead{$b$} & \colhead{$\sigma(b)$}}
\startdata
\cutinhead{$Z = 0.0005$}
$0.934	$ & $0.013	$ & $-0.826$ & $	0.012$ & $	-1.902$ & $	0.045$ \\
$0.877	$ & $0.018	$ & $-0.815$ & $	0.007$ & $	-1.890$ & $	0.024$ \\
$0.776	$ & $0.022	$ & $-0.800$ & $	0.007$ & $	-1.861$ & $	0.020$ \\
$0.627	$ & $0.027	$ & $-0.786$ & $	0.006$ & $	-1.825$ & $	0.017$ \\
$0.414	$ & $0.028	$ & $-0.772$ & $	0.005$ & $	-1.788$ & $	0.015$ \\
$0.167	$ & $0.029	$ & $-0.760$ & $	0.005$ & $	-1.757$ & $	0.015$ \\
$-0.102$ & $	0.031$ & $	-0.754$ & $	0.004$ & $	-1.739$ & $	0.012$ \\
$-0.358$ & $	0.031$ & $	-0.750$ & $	0.003$ & $	-1.727$ & $	0.011$ \\
$-0.590$ & $	0.025$ & $	-0.748$ & $	0.004$ & $	-1.717$ & $	0.014$ \\
$-0.765$ & $	0.021$ & $	-0.747$ & $	0.004$ & $	-1.710$ & $	0.018$ \\
$-0.883$ & $	0.014$ & $	-0.746$ & $	0.005$ & $	-1.706$ & $	0.021$ \\
$-0.950$ & $	0.010$ & $	-0.747$ & $	0.008$ & $	-1.709$ & $	0.040$ \\
\cutinhead{$Z = 0.0010$}
$0.940	$ & $0.011	$ & $-0.795$ & $	0.020$ & $	-1.981$ & $	0.057$ \\
$0.873	$ & $0.018	$ & $-0.774$ & $	0.014$ & $	-1.934$ & $	0.039$ \\
$0.744	$ & $0.025	$ & $-0.751$ & $	0.010$ & $	-1.879$ & $	0.030$ \\
$0.556	$ & $0.034	$ & $-0.733$ & $	0.008$ & $	-1.830$ & $	0.024$ \\
$0.282	$ & $0.033	$ & $-0.717$ & $	0.007$ & $	-1.786$ & $	0.021$ \\
$-0.037$ & $	0.035$ & $	-0.705$ & $	0.007$ & $	-1.752$ & $	0.020$ \\
$-0.342$ & $	0.036$ & $	-0.696$ & $	0.006$ & $	-1.726$ & $	0.019$ \\
$-0.603$ & $	0.025$ & $	-0.691$ & $	0.005$ & $	-1.710$ & $	0.017$ \\
$-0.789$ & $	0.022$ & $	-0.687$ & $	0.006$ & $	-1.698$ & $	0.020$ \\
$-0.906$ & $	0.015$ & $	-0.685$ & $	0.008$ & $	-1.689$ & $	0.030$ \\
$-0.963$ & $	0.008$ & $	-0.684$ & $	0.011$ & $	-1.685$ & $	0.049$ \\
\cutinhead{$Z = 0.0020$}
$0.965	$ & $0.009	$ & $-0.748$ & $	0.042$ & $	-2.005$ & $	0.123$ \\
$0.910	$ & $0.015	$ & $-0.728$ & $	0.027$ & $	-1.955$ & $	0.072$ \\
$0.794	$ & $0.025	$ & $-0.702$ & $	0.020$ & $	-1.890$ & $	0.056$ \\
$0.594	$ & $0.029	$ & $-0.681$ & $	0.013$ & $	-1.837$ & $	0.038$ \\
$0.307	$ & $0.036	$ & $-0.667$ & $	0.012$ & $	-1.800$ & $	0.034$ \\
$-0.023$ & $	0.034$ & $	-0.653$ & $	0.010$ & $	-1.763$ & $	0.028$ \\
$-0.356$ & $	0.035$ & $	-0.647$ & $	0.008$ & $	-1.745$ & $	0.024$ \\
$-0.630$ & $	0.032$ & $	-0.638$ & $	0.009$ & $	-1.721$ & $	0.027$ \\
$-0.822$ & $	0.021$ & $	-0.635$ & $	0.011$ & $	-1.713$ & $	0.034$ \\
$-0.928$ & $	0.012$ & $	-0.628$ & $	0.014$ & $	-1.691$ & $	0.044$ \\
$-0.974$ & $	0.008$ & $	-0.625$ & $	0.020$ & $	-1.682$ & $	0.076$ \\
\cutinhead{$Z = 0.0060$}
$0.922	$ & $0.015	$ & $-0.649$ & $	0.052$ & $	-1.989$ & $	0.125$ \\
$0.810	$ & $0.021	$ & $-0.623$ & $	0.033$ & $	-1.929$ & $	0.080$ \\
$0.601	$ & $0.034	$ & $-0.606$ & $	0.024$ & $	-1.891$ & $	0.059$ \\
$0.298	$ & $0.039	$ & $-0.591$ & $	0.019$ & $	-1.860$ & $	0.049$ \\
$-0.070$ & $	0.039$ & $	-0.575$ & $	0.019$ & $	-1.826$ & $	0.047$ \\
$-0.420$ & $	0.036$ & $	-0.563$ & $	0.017$ & $	-1.800$ & $	0.044$ \\
$-0.693$ & $	0.025$ & $	-0.553$ & $	0.021$ & $	-1.777$ & $	0.053$ \\
$-0.868$ & $	0.018$ & $	-0.536$ & $	0.023$ & $	-1.739$ & $	0.062$ \\
$-0.951$ & $	0.012$ & $	-0.534$ & $	0.033$ & $	-1.733$ & $	0.092$ \\
\hline
\enddata
\end{deluxetable}

\begin{equation}
M_X = a + b \, \log P,
\end{equation}

\begin{deluxetable}{cccccc}  
\tabletypesize{\footnotesize}
\tablecaption{RRL PL Relation in $H$: Coefficients of the Fits} 
\tablewidth{0pt}
\tablehead{
\colhead{$\LeeZinn$} & \colhead{$\sigma({\LeeZinn})$} & \colhead{$a$} & 
\colhead{$\sigma(a)$} & \colhead{$b$} & \colhead{$\sigma(b)$}}
\startdata
\cutinhead{$Z = 0.0005$}
$0.934	$ & $0.013	$ & $-1.136$ & $	0.003$ & $	-2.311$ & $	0.013$ \\
$0.877	$ & $0.018	$ & $-1.137$ & $	0.002$ & $	-2.315$ & $	0.009$ \\
$0.776	$ & $0.022	$ & $-1.137$ & $	0.002$ & $	-2.317$ & $	0.008$ \\
$0.627	$ & $0.027	$ & $-1.137$ & $	0.002$ & $	-2.316$ & $	0.007$ \\
$0.414	$ & $0.028	$ & $-1.137$ & $	0.002$ & $	-2.316$ & $	0.006$ \\
$0.167	$ & $0.029	$ & $-1.138$ & $	0.002$ & $	-2.315$ & $	0.007$ \\
$-0.102$ & $	0.031$ & $	-1.139$ & $	0.002$ & $	-2.316$ & $	0.006$ \\
$-0.358$ & $	0.031$ & $	-1.141$ & $	0.002$ & $	-2.320$ & $	0.006$ \\
$-0.590$ & $	0.025$ & $	-1.142$ & $	0.002$ & $	-2.320$ & $	0.008$ \\
$-0.765$ & $	0.021$ & $	-1.143$ & $	0.003$ & $	-2.323$ & $	0.011$ \\
$-0.883$ & $	0.014$ & $	-1.144$ & $	0.003$ & $	-2.322$ & $	0.013$ \\
$-0.950$ & $	0.010$ & $	-1.146$ & $	0.005$ & $	-2.327$ & $	0.027$ \\
\cutinhead{$Z = 0.0010$}
$0.940	$ & $0.011	$ & $-1.101$ & $	0.006$ & $	-2.358$ & $	0.019$ \\
$0.873	$ & $0.018	$ & $-1.096$ & $	0.005$ & $	-2.346$ & $	0.014$ \\
$0.744	$ & $0.025	$ & $-1.090$ & $	0.003$ & $	-2.332$ & $	0.009$ \\
$0.556	$ & $0.034	$ & $-1.086$ & $	0.003$ & $	-2.322$ & $	0.009$ \\
$0.282	$ & $0.033	$ & $-1.083$ & $	0.002$ & $	-2.310$ & $	0.007$ \\
$-0.037$ & $	0.035$ & $	-1.081$ & $	0.002$ & $	-2.304$ & $	0.008$ \\
$-0.342$ & $	0.036$ & $	-1.079$ & $	0.003$ & $	-2.297$ & $	0.008$ \\
$-0.603$ & $	0.025$ & $	-1.079$ & $	0.002$ & $	-2.295$ & $	0.008$ \\
$-0.789$ & $	0.022$ & $	-1.079$ & $	0.003$ & $	-2.295$ & $	0.011$ \\
$-0.906$ & $	0.015$ & $	-1.080$ & $	0.004$ & $	-2.292$ & $	0.016$ \\
$-0.963$ & $	0.008$ & $	-1.080$ & $	0.006$ & $	-2.292$ & $	0.025$ \\
\cutinhead{$Z = 0.0020$}
$0.965	$ & $0.009	$ & $-1.058$ & $	0.015$ & $	-2.380$ & $	0.045$ \\
$0.910	$ & $0.015	$ & $-1.051$ & $	0.009$ & $	-2.361$ & $	0.025$ \\
$0.794	$ & $0.025	$ & $-1.042$ & $	0.007$ & $	-2.339$ & $	0.021$ \\
$0.594	$ & $0.029	$ & $-1.035$ & $	0.005$ & $	-2.321$ & $	0.014$ \\
$0.307	$ & $0.036	$ & $-1.031$ & $	0.005$ & $	-2.308$ & $	0.013$ \\
$-0.023$ & $	0.034$ & $	-1.027$ & $	0.004$ & $	-2.297$ & $	0.012$ \\
$-0.356$ & $	0.035$ & $	-1.025$ & $	0.003$ & $	-2.291$ & $	0.010$ \\
$-0.630$ & $	0.032$ & $	-1.023$ & $	0.004$ & $	-2.284$ & $	0.012$ \\
$-0.822$ & $	0.021$ & $	-1.023$ & $	0.004$ & $	-2.282$ & $	0.014$ \\
$-0.928$ & $	0.012$ & $	-1.021$ & $	0.006$ & $	-2.275$ & $	0.019$ \\
$-0.974$ & $	0.008$ & $	-1.020$ & $	0.009$ & $	-2.271$ & $	0.034$ \\
\cutinhead{$Z = 0.0060$}
$0.922	$ & $0.015	$ & $-0.975$ & $	0.022$ & $	-2.381$ & $	0.053$ \\
$0.810	$ & $0.021	$ & $-0.964$ & $	0.014$ & $	-2.355$ & $	0.035$ \\
$0.601	$ & $0.034	$ & $-0.957$ & $	0.010$ & $	-2.338$ & $	0.026$ \\
$0.298	$ & $0.039	$ & $-0.950$ & $	0.008$ & $	-2.325$ & $	0.021$ \\
$-0.070$ & $	0.039$ & $	-0.944$ & $	0.008$ & $	-2.312$ & $	0.020$ \\
$-0.420$ & $	0.036$ & $	-0.939$ & $	0.007$ & $	-2.301$ & $	0.019$ \\
$-0.693$ & $	0.025$ & $	-0.935$ & $	0.009$ & $	-2.292$ & $	0.023$ \\
$-0.868$ & $	0.018$ & $	-0.929$ & $	0.010$ & $	-2.277$ & $	0.027$ \\
$-0.951$ & $	0.012$ & $	-0.928$ & $	0.014$ & $	-2.274$ & $	0.040$ \\
\hline
\enddata
\end{deluxetable}

\noindent with

\begin{equation}
a = \sum_{i = 0}^{3} a_{i} \left(\LeeZinn\right)^{i}, \,\,\,\, 
b = \sum_{i = 0}^{3} b_{i} \left(\LeeZinn\right)^{i}. 
\end{equation} 

\noindent For all of the $VRIJHK$ passbands, 
the $a_i$, $b_i$ coefficients are provided in Table~9.

\begin{deluxetable}{cccccc}  
\tabletypesize{\footnotesize}
\tablecaption{RRL PL Relation in $K$: Coefficients of the Fits} 
\tablewidth{0pt}
\tablehead{
\colhead{$\LeeZinn$} & \colhead{$\sigma({\LeeZinn})$} & \colhead{$a$} & 
\colhead{$\sigma(a)$} & \colhead{$b$} & \colhead{$\sigma(b)$}}
\startdata
\cutinhead{$Z = 0.0005$}
$0.934	$ & $0.013	$ & $-1.168$ & $	0.002$ & $	-2.343$ & $	0.012$ \\
$0.877	$ & $0.018	$ & $-1.169$ & $	0.002$ & $	-2.348$ & $	0.009$ \\
$0.776	$ & $0.022	$ & $-1.170$ & $	0.002$ & $	-2.352$ & $	0.007$ \\
$0.627	$ & $0.027	$ & $-1.171$ & $	0.002$ & $	-2.352$ & $	0.007$ \\
$0.414	$ & $0.028	$ & $-1.172$ & $	0.002$ & $	-2.355$ & $	0.006$ \\
$0.167	$ & $0.029	$ & $-1.173$ & $	0.002$ & $	-2.355$ & $	0.007$ \\
$-0.102$ & $	0.031$ & $	-1.175$ & $	0.002$ & $	-2.358$ & $	0.006$ \\
$-0.358$ & $ 	0.031$ & $	-1.177$ & $	0.002$ & $	-2.362$ & $	0.006$ \\
$-0.590$ & $	0.025$ & $	-1.178$ & $	0.002$ & $	-2.364$ & $	0.008$ \\
$-0.765$ & $	0.021$ & $	-1.180$ & $	0.003$ & $	-2.368$ & $	0.011$ \\
$-0.883$ & $	0.014$ & $	-1.181$ & $	0.003$ & $	-2.367$ & $	0.013$ \\
$-0.950$ & $	0.010$ & $	-1.183$ & $	0.005$ & $	-2.373$ & $	0.026$ \\
\cutinhead{$Z = 0.0010$}
$0.940	$ & $0.011	$ & $-1.133$ & $	0.005$ & $	-2.388$ & $	0.017$ \\
$0.873	$ & $0.018	$ & $-1.128$ & $	0.004$ & $	-2.379$ & $	0.013$ \\
$0.744	$ & $0.025	$ & $-1.124$ & $	0.003$ & $	-2.367$ & $	0.009$ \\
$0.556	$ & $0.034	$ & $-1.121$ & $	0.003$ & $	-2.359$ & $	0.009$ \\
$0.282	$ & $0.033	$ & $-1.118$ & $	0.002$ & $	-2.350$ & $	0.007$ \\
$-0.037$ & $	0.035$ & $	-1.11$ & $7	0.002$ & $	-2.345$ & $	0.008$ \\
$-0.342$ & $	0.036$ & $	-1.11$ & $5	0.002$ & $	-2.339$ & $	0.008$ \\
$-0.603$ & $	0.025$ & $	-1.11$ & $6	0.002$ & $	-2.338$ & $	0.008$ \\
$-0.789$ & $	0.022$ & $	-1.11$ & $7	0.003$ & $	-2.339$ & $	0.011$ \\
$-0.906$ & $	0.015$ & $	-1.11$ & $7	0.004$ & $	-2.337$ & $	0.016$ \\
$-0.963$ & $	0.008$ & $	-1.11$ & $7	0.006$ & $	-2.338$ & $	0.024$ \\
\cutinhead{$Z = 0.0020$}
$0.965	$ & $0.009	$ & $-1.091$ & $	0.014$ & $	-2.410$ & $	0.041$ \\
$0.910	$ & $0.015	$ & $-1.084$ & $	0.008$ & $	-2.393$ & $	0.022$ \\
$0.794	$ & $0.025	$ & $-1.077$ & $	0.007$ & $	-2.374$ & $	0.019$ \\
$0.594	$ & $0.029	$ & $-1.071$ & $	0.004$ & $	-2.358$ & $	0.013$ \\
$0.307	$ & $0.036	$ & $-1.067$ & $	0.004$ & $	-2.347$ & $	0.012$ \\
$-0.023$ & $	0.034$ & $	-1.064$ & $	0.004$ & $	-2.337$ & $	0.011$ \\
$-0.356$ & $	0.035$ & $	-1.063$ & $	0.003$ & $	-2.333$ & $	0.010$ \\
$-0.630$ & $	0.032$ & $	-1.061$ & $	0.004$ & $	-2.327$ & $	0.012$ \\
$-0.822$ & $	0.021$ & $	-1.061$ & $	0.004$ & $	-2.325$ & $	0.013$ \\
$-0.928$ & $	0.012$ & $	-1.059$ & $	0.006$ & $	-2.320$ & $	0.018$ \\
$-0.974$ & $	0.008$ & $	-1.059$ & $	0.009$ & $	-2.317$ & $	0.032$ \\
\cutinhead{$Z = 0.0060$}
$0.922	$ & $0.015	$ & $-1.011$ & $	0.021$ & $	-2.413$ & $	0.049$ \\
$0.810	$ & $0.021	$ & $-1.001$ & $	0.013$ & $	-2.389$ & $	0.032$ \\
$0.601	$ & $0.034	$ & $-0.994$ & $	0.010$ & $	-2.374$ & $	0.024$ \\
$0.298	$ & $0.039	$ & $-0.988$ & $	0.008$ & $	-2.362$ & $	0.019$ \\
$-0.070$ & $	0.039$ & $	-0.983$ & $	0.008$ & $	-2.349$ & $	0.019$ \\
$-0.420$ & $	0.036$ & $	-0.978$ & $	0.007$ & $	-2.340$ & $	0.018$ \\
$-0.693$ & $	0.025$ & $	-0.974$ & $	0.008$ & $	-2.331$ & $	0.021$ \\
$-0.868$ & $	0.018$ & $	-0.968$ & $	0.009$ & $	-2.317$ & $	0.025$ \\
$-0.951$ & $	0.012$ & $	-0.968$ & $	0.013$ & $	-2.315$ & $	0.037$ \\
\hline
\enddata
\end{deluxetable}

\section{Remarks on the RRL PL Relations} 
Figures~3 and 4 reveal a complex pattern for the variation of the coefficients
of the PL relation as a function of HB morphology. While, as anticipated, 
the dependence on HB type (particularly the slope) is quite small for 
the redder passbands (note the much smaller axis scale range for the 
corresponding $H$ and $K$ plots than for the remaining ones), the same 
cannot be said with respect to the bluer passbands, particularly $U$ 
and $B$, for which one does see marked variations as one moves from 
very red to very blue HB types. This is obviously due to the much more 
important effects of evolution away from the zero-age HB in the bluer 
passbands. In order to fully highlight the changes in the PL relations 
in each of the considered bandpasses, we show, in Figures~5 through 12, 
the changes 
in the absolute magnitude--log-period distributions for each bandpass, 
from $U$ (Fig.~5) to $K$ (Fig.~12), for a representative metallicity, 
$Z = 0.001$. Each figure is comprised of a mosaic of 10 plots, each for 
a different HB type, from very red (upper left panels) to very blue 
(lower right panels). In the bluer passbands, one can see the stars 
that are evolved away from a position on the blue zero-age HB (and thus 
brighter for a given period) gradually becoming more dominant as the HB 
type gets bluer. As already discussed, the effects of luminosity 
and temperature upon the period-absolute magnitude distribution are 
almost orthogonal in these bluer passbands. As a consequence, when the 
number of stars evolved away from the 
blue zero-age HB becomes comparable to the number of stars on the main 
phase of the HB, which occurs at $\LeeZinn \sim 0.4-0.8$, a sharp break 
in slope results, for the $U$ and $B$ passbands, at around these HB types. 
The effect is more pronounced at the lower metallicities, where the 
evolutionary effect is expected to be more important (e.g., Catelan 1993). 
For the redder passbands, including the visual, the changes are smoother 
as a function of HB morphology.

\begin{deluxetable*}{ccccccccccc}  
\tabletypesize{\footnotesize}
\tablecaption{Coefficients of the RRL PL Relation in $BVRIJHK$: Analytical Fits} 
\tablewidth{0pt}
\tablehead{ 
\colhead{Metallicity} &  \colhead{} &
\colhead{$a_{0}$} & \colhead{$a_{1}$} & \colhead{$a_{2}$} & \colhead{$a_{3}$} & \colhead{} &  
\colhead{$b_{0}$} & \colhead{$b_{1}$} & \colhead{$b_{2}$} & \colhead{$b_{3}$}}
\startdata
\cutinhead{$V$}
$Z = 0.0060$ & & $0.5276$ & $-0.0420$ & $0.1056$ &  $-0.2011$ && $-0.9440$ & $-0.0436$ & $0.3207$ & $-0.5303$ \\
$Z = 0.0020$ & & $0.4470$ & $-0.0232$ & $0.0666$ &  $-0.2309$ && $-0.8512$ & $0.0122$ & $0.3107$ & $-0.7248$ \\
$Z = 0.0010$ & & $0.3976$ & $-0.0416$ & $0.0482$ &  $-0.2117$ && $-0.8243$ & $-0.0231$ & $0.3212$ & $-0.7169$ \\
$Z = 0.0005$ & & $0.3668$ & $-0.0314$ & $-0.0053$ & $-0.1550$ && $-0.7455$ & $0.0407$ & $0.1342$ & $-0.4966$  \\
\cutinhead{$R$}
$Z = 0.0060$ & & $0.2107$ & $-0.0701$ & $-0.0204$ & $-0.0764$& & $-1.0050$ & $-0.1507$ & $-0.0390$ & $-0.1723$ \\
$Z = 0.0020$ & & $0.1469$ & $-0.0566$ & $-0.0633$ & $-0.0995$& & $-0.8523$ & $-0.1500$ & $-0.1519$ & $-0.2521$ \\
$Z = 0.0010$ & & $0.1020$ & $-0.0808$ & $-0.0784$ & $-0.0744$& & $-0.8086$ & $-0.2397$ & $-0.1809$ & $-0.1608$ \\
$Z = 0.0005$ & & $0.0659$ & $-0.0811$ & $-0.0832$ & $-0.0537$& & $-0.7553$ & $-0.2482$ & $-0.1769$ & $-0.0677$ \\
\cutinhead{$I$}
$Z = 0.0060$ & & $-0.0162$ & $-0.0540$ & $-0.0216$ & $-0.0640$& & $-1.2459$ & $-0.1155$ & $-0.0442$ & $-0.1466$ \\
$Z = 0.0020$ & & $-0.0913$ & $-0.0393$ & $-0.0568$ & $-0.0788$& & $-1.1499$ & $-0.1058$ & $-0.1397$ & $-0.2002$ \\
$Z = 0.0010$ & & $-0.1445$ & $-0.0545$ & $-0.0680$ & $-0.0614$& & $-1.1340$ & $-0.1647$ & $-0.1614$ & $-0.1378$ \\
$Z = 0.0005$ & & $-0.1939$ & $-0.0515$ & $-0.0679$ & $-0.0450$ & & $-1.1192$ & $-0.1640$ & $-0.1474$ & $-0.0655$ \\
\cutinhead{$J$}
$Z = 0.0060$ & & $-0.5766$ & $-0.0278$ & $-0.0150$ & $-0.0378$ & & $-1.8291$ & $-0.0581$ & $-0.0318$ & $-0.0874$ \\
$Z = 0.0020$ & & $-0.6517$ & $-0.0181$ & $-0.0335$ & $-0.0452$ & & $-1.7599$ & $-0.0498$ & $-0.0809$ & $-0.1167$ \\
$Z = 0.0010$ & & $-0.7023$ & $-0.0250$ & $-0.0383$ & $-0.0355$ & & $-1.7478$ & $-0.0786$ & $-0.0891$ & $-0.0824$ \\
$Z = 0.0005$ & & $-0.7546$ & $-0.0209$ & $-0.0343$ & $-0.0238$ & & $-1.7438$ & $-0.0728$ & $-0.0706$ & $-0.0379$ \\
\cutinhead{$H$}
$Z = 0.0060$ & & $-0.9447$ & $-0.0110$ & $-0.0070$ & $-0.0160$ & & $-2.3128$ & $-0.0231$ & $-0.0149$ & $-0.0375$ \\
$Z = 0.0020$ & & $-1.0262$ & $-0.0041$ & $-0.0124$ & $-0.0152$ & & $-2.2953$ & $-0.0138$ & $-0.0292$ & $-0.0414$ \\
$Z = 0.0010$ & & $-1.0798$ & $-0.0037$ & $-0.0108$ & $-0.0079$ & & $-2.3019$ & $-0.0170$ & $-0.0242$ & $-0.0186$ \\
$Z = 0.0005$ & & $-1.1385$ & $0.0037$ & $-0.0027$ & $0.0012$ & & $-2.3169$ & $0.0016$ & $-0.0029$ & $0.0056$ \\
\cutinhead{$K$}
$Z = 0.0060$ & & $-0.9829$ & $-0.0101$ & $-0.0066$ & $-0.0148$ & & $-2.3499$ & $-0.0210$ & $-0.0140$ & $-0.0346$ \\
$Z = 0.0020$ & & $-1.0630$ & $-0.0032$ & $-0.0113$ & $-0.0132$ & & $-2.3357$ & $-0.0115$ & $-0.0268$ & $-0.0361$ \\
$Z = 0.0010$ & & $-1.1159$ & $-0.0024$ & $-0.0093$ & $-0.0061$ & & $-2.3427$ & $-0.0131$ & $-0.0212$ & $-0.0142$ \\
$Z = 0.0005$ & & $-1.1742$ & $0.0051$ & $-0.0012$ & $0.0028$ & & $-2.3580$ & $0.0061$ & $-0.0004$ & $0.0090$ \\
\hline 
\enddata
\end{deluxetable*}

The dependence of the RRL PL relation in $IJHK$ on the adopted width of 
the mass distribution, as well as on the helium abundance, has been 
analyzed by computing additional sets of synthetic HBs for 
$\sigma_M = 0.030 \, M_{\odot}$ ($Z = 0.001$) and for a main-sequence 
helium abundance of 28\% ($Z = 0.002$). The results are shown in Figure~13. 
As can be seen from the $I$ plots, the precise shape of the mass distribution 
plays but a minor role in defining the PL relation. On the other hand, the 
effects of a significantly enhanced helium abundance can be much more 
important, particularly in regard to the zero point of the PL relations 
in all four passbands. Therefore, caution is recommended when employing 
locally calibrated RRL PL relations to extragalactic environments, in view 
of the possibility of different chemical enrichment laws. 
From a theoretical point of view, a conclusive assessment of the effects 
of helium diffusion on the main sequence, dredge-up on the first ascent 
of the RGB, and non-canonical helium mixing on the upper RGB, will 
all be required before calibrations such as the present ones can be 
considered final.

\section{``Average'' Relations} 
In applications of the RRL PL relations presented in this paper thus 
far to derive distances to objects whose HB types are not known a priori, 
as may easily happen in the case of distant galaxies for instance, some 
``average'' form of the PL relation might be useful which does not 
explicitly show a dependence on HB morphology. In the redder passbands, 
in particular, a meaningful relation of that type may be obtained when one 
considers that the dependence of the zero points and slopes of the 
corresponding relations, as presented in the previous sections, is 
fairly mild. Therefore, in the present section, we present ``average'' 
relations for $I$, $J$, $H$, $K$, obtaining by simply gathering together 
the 389,484 stars in all of the simulations for all HB types and 
metallicities ($0.0005 \leq Z \leq 0.006$). 
Utilizing a simple least-squares procedure with the 
log-periods and log-metallicities as independent variables, we obtain 
the following fits:\footnote{For all equations presented in this section, 
the statistical errors in the derived coefficients are always very small, 
of order $10^{-5}-10^{-3}$, due to the very large number of stars involved 
in the corresponding fits. Consequently, we omit them from the equations 
that we provide. Undoubtedly, the main sources of error affecting 
these relations are systematic rather than statistical---e.g., helium 
abundances (see \S5), bolometric corrections, temperature coefficient 
of the period-mean density relation, etc..} 

\begin{equation}
M_I = 0.4711   - 1.1318   \, \log P + 0.2053  \, \log Z, 
\end{equation} 

\noindent with a correlation coefficient $r = 0.967$; 

\begin{equation}
M_J = -0.1409  - 1.7734   \, \log P + 0.1899  \, \log Z, 
\end{equation} 

\noindent with a correlation coefficient $r = 0.9936$; 

\begin{equation}
M_H = -0.5508  - 2.3134   \, \log P + 0.1780  \, \log Z, 
\end{equation} 
 
\noindent with a correlation coefficient $r = 0.9991$; 

\begin{equation}
M_K = -0.5968  - 2.3529   \, \log P + 0.1746  \, \log Z,   
\end{equation} 

\noindent with a correlation coefficient $r = 0.9992$. Note that the latter 
relation is of the same form as the one presented by Bono et al. (2001). The 
metallicity dependence we derive is basically identical to that in Bono et 
al., whereas the $\log P$ slope is slightly steeper (by 0.28), in absolute 
value, in our case. In terms of zero points, the two relations, at 
representative values $P = 0.50$~d and $Z = 0.001$, provide $K$-band 
magnitudes which differ by only 0.05~mag, ours being slightly brighter.

\begin{figure*}[ht]
  \figurenum{13}
  \epsscale{0.9}
\plotone{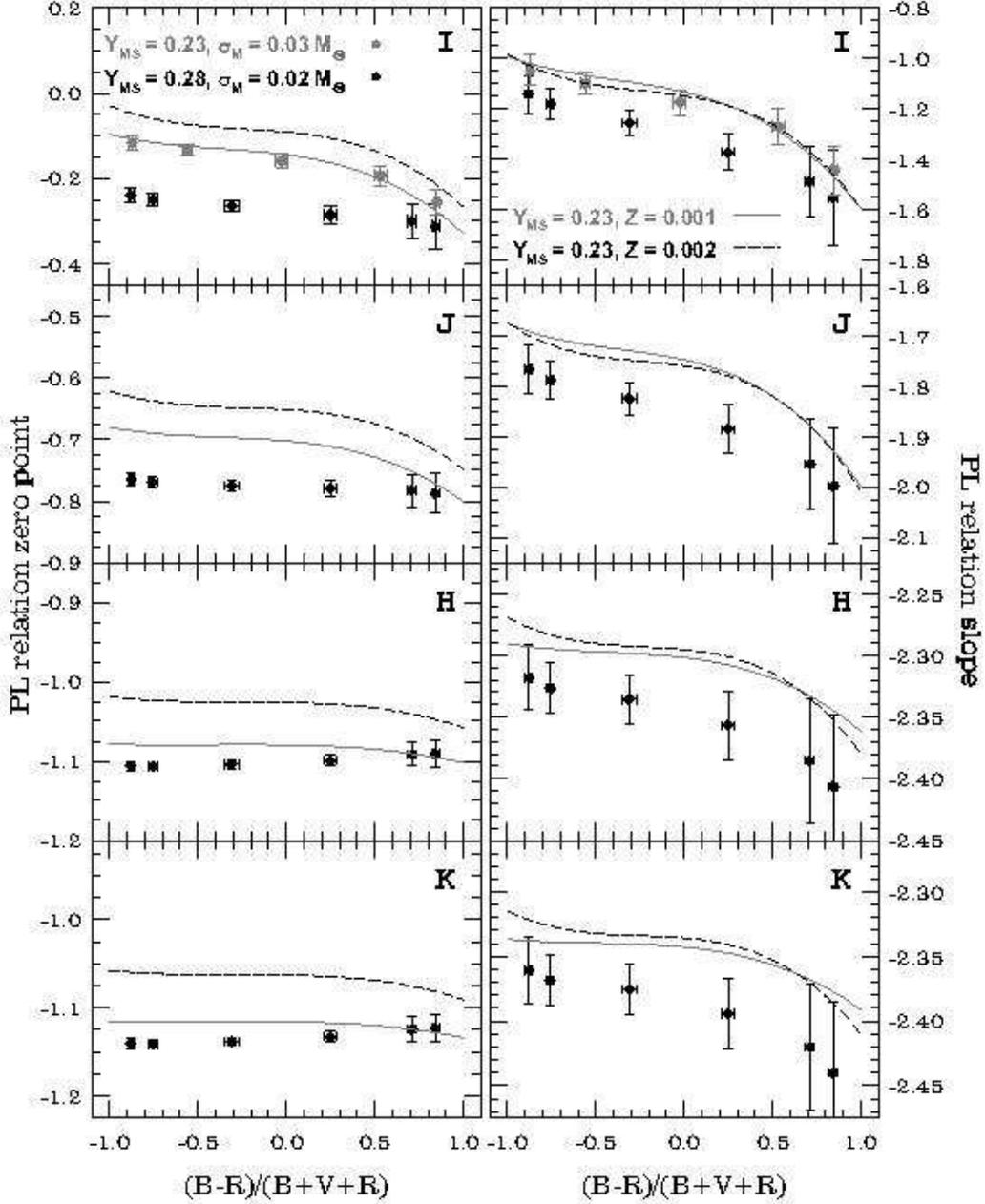}
  \caption{Effects of $\sigma_M$ and of the helium abundance upon the 
     RRL PL relation in the $IJHK$ bands. 
     The lines indicate the analytical fits obtained, from 
     equations~(1) and (2), for our assumed case of $Y_{\rm MS} = 0.23$ 
     and $\sigma_M = 0.02\, M_{\odot}$ for a metallicity $Z = 0.001$ 
     (thick gray lines) or $Z = 0.002$ (dashed lines).
     In the two upper panels, the results of 
     an additional set of models, computed by increasing the mass 
     dispersion from $0.02\,M_{\odot}$ to $0.03\,M_{\odot}$, are 
     shown (gray circles). As one can see, $\sigma_M$ does not 
     affect the relation in a significant way, so that similar 
     results for an increased $\sigma_M$ are omitted in the lower 
     panels. The helium abundance, in turn, is seen to play a much 
     more important role, particularly in defining the zero point 
     of the PL relation. 
        }
      \label{Fig13}
\end{figure*}

In addition, the same type of exercise can provide us with an average 
relation between HB magnitude in the visual and metallicity. Performing 
an ordinary least-squares fit of the form $M_V = f(\log Z)$ (i.e., 
with $\log Z$ as the independent variable), we obtain: 

\begin{equation}
M_V = 1.455  + 0.277  \, \log Z,   
\end{equation} 

\noindent with a correlation coefficient $r = 0.83$. 

The above equation has a stronger slope than is often adopted in the 
literature (e.g., Chaboyer 1999). This is likely due to the fact that 
the $M_V - {\rm [M/H]}$ relation is actually non-linear, with the slope increasing 
for $Z \gtrsim 0.001$ (Castellani et al. 1991), where most of our simulations 
will be found. A quadratic version of the same equation reads as follows: 

\begin{equation}
M_V = 2.288  + 0.8824  \, \log Z  + 0.1079 \, (\log Z)^{2}.  
\end{equation} 

\noindent As one can see, the slope provided by this relation, at a 
metallicity $Z = 0.001$, is 0.235, thus fully compatible with the 
range discussed by Chaboyer (1999). 

The last two equations can also be placed in their more usual form, with 
[M/H] (or [Fe/H]) as the independent variable, if we recall, from Sweigart \& 
Catelan (1998), that the solar metallicity corresponding to the (scaled-solar) 
evolutionary models 
utilized in the present paper is $Z = 0.01716$. Therefore, the conversion 
between $Z$ and [M/H] that is appropriate for our models is as follows: 

\begin{equation}
\log Z = {\rm [M/H]} - 1.765. 
\end{equation} 

\noindent The effects of an enhancement in $\alpha$-capture elements 
with respect to a solar-scaled mixture, as indeed observed among most metal-poor 
stars in the Galactic halo, can be taken into account by the following scaling 
relation (Salaris, Chieffi, \& Straniero 1993): 

\begin{equation}
{\rm [M/H]} = {\rm [Fe/H]} + \log (0.638\,f + 0.362), 
\end{equation} 

\noindent where $f = 10^{\rm [\alpha/Fe]}$. Note that such a relation should 
be used with due care for metallicities $Z > 0.003$ (VandenBerg et al. 2000). 

With these equations in mind, the linear version of the $M_V - {\rm [M/H]}$ 
relation becomes 

\begin{equation}
M_V = 0.967 + 0.277 \, {\rm [M/H]}, 
\end{equation} 

\noindent whereas the quadratic one reads instead

\begin{equation}
M_V = 1.067 + 0.502 \, {\rm [M/H]} + 0.108 \, {\rm [M/H]}^{2}. 
\end{equation} 

\noindent The latter equation provides $M_V = 0.60$~mag at ${\rm [Fe/H]} = -1.5$ 
(assuming ${\rm [\alpha/Fe]} \simeq 0.3$; e.g., Carney 1996), 
in very good agreement with the favored values in Chaboyer (1999) and Cacciari 
(2003)---thus supporting a distance modulus for the LMC of $(m-M)_0 = 18.47$~mag.

\begin{figure*}[ht]
  \figurenum{14}
  \epsscale{0.65}
\plotone{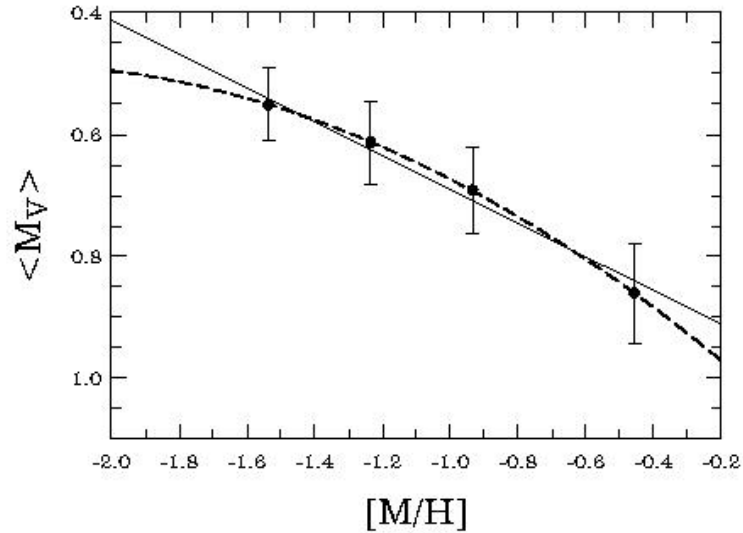}
  \caption{Predicted correlation between average RRL $V$-band absolute 
     magnitude and metallicity. Each of the four datapoints represents 
     the average magnitude over the full sample of simulations for that 
     metallicity. The ``error bars'' actually indicate the standard 
     deviation of the mean. The full and dashed lines show equations 
     (11) and (12), respectively.    
        }
      \label{Fig14}
\end{figure*}


To close, we note that the present models, which cover only a modest range in 
metallicities, do not provide useful input regarding the question of whether 
the $M_V - {\rm [M/H]}$ relation is better described by a parabola or by two 
straight lines (Bono et al. 2003). To illustrate this, we show, in Figure~14, 
the average RRL magnitudes for each metallicity value considered, along with 
equations~(11) and (12). Note that the ``error bars'' actually represent the 
standard deviation of the mean over the full set of RRL stars in the simulations 
for each [M/H] value.

\section{Conclusions} 
We have presented RRL PL relations in the bandpasses of the 
Johnsons-Cousins-Glass $UBVRIJHK$ system. While in the case 
of the Cepheids the existence of a PL relation is a necessary 
consequence of the large range in luminosities encompassed by these 
variables, in the case of RRL stars useful PL relations are instead 
primarily due to the occasional presence of a large range in bolometric 
corrections when going from the blue to the red edge of the RRL instability 
strip. This leads to particularly useful PL relations in $IJHK$, where 
the effects of luminosity and temperature conspire to produce tight 
relations. In bluer passbands, on the other hand, the effects of 
luminosity do not reinforce those of temperature, leading to the 
presence of large scatter in the relations and to a stronger dependence 
on evolutionary effects. We provide a detailed tabulation of our 
derived slopes and zero points for four different metallicities and 
covering virtually the whole range in HB morphology, from very red to 
very blue, fully taking into account, for the first time, the detailed 
effects of evolution away from the zero-age HB upon the derived PL 
relations in all of the $UBVRIJHK$ passbands. We also provide ``average'' 
PL relations in $IJHK$, for applications in cases where the HB type is 
not known a priori; as well as a new calibration of the 
$M_V - {\rm [Fe/H]}$ relation.  
In Paper~II, we 
will provide comparisons between these results and the observations, 
particularly in $I$, where we expect our new calibration to be 
especially useful due to the wide availability and ease of 
observations in this filter, in comparison with $JHK$.

\acknowledgments M.C. acknowledges support by Proyecto FONDECYT Regular 
No. 1030954. B.J.P. would like to thank the National Science Foundation 
(NSF) for support through a CAREER award, AST 99-84073. H.A.S. acknowledges 
the NSF for support under grant AST 02-05813.


\begin{references}
\reference{} Bono, G., Caputo, F., Castellani, V., Marconi, M., \& Storm, J.,
2001, \mnras, 326, 1183
\reference{} Bono, G., Caputo, F., Castellani, V., Marconi, M., Storm, J., \&
Degl'Innocenti, S. 2003, \mnras, 344, 1097
\reference{} Cacciari, C. 2003, in New Horizons in Globular Cluster Astronomy, 
   ASP Conf. Ser., Vol. 296, 
   ed. G. Piotto, G. Meylan, S. G. Djorgovski, \& M. Riello (San Francisco: 
   ASP), 329 
\reference{} Caputo, F., De Stefanis, P., Paez, E., \& Quarta, M. L. 1987, \aaps, 
  68, 119 
\reference{} Caputo, F., Marconi, M., \& Santolamazza, P. 1998, \mnras, 293, 364
\reference{} Carney, B. W. 1996, \pasp, 108, 900
\reference{} Castellani, V., Chieffi, A., \& Pulone, L. 1991, \apjs, 76, 911
\reference{} Catelan, M. 1993, \aaps, 98, 547
\reference{} Catelan, M. 2004a, \apj, 600, 409
\reference{} Catelan, M. 2004b, in Variable Stars in the Local Group, ASP Conf. 
   Ser., Vol. 310, 
   ed. D. W. Kurtz \& K. Pollard (San Francisco: ASP), in press (astro-ph/0310159) 
\reference{} Catelan, M., Borissova, J., Sweigart, A.~V., \& Spassova, N.\ 1998, 
  \apj, 494, 265 
\reference{} Chaboyer, B. 1999, in Post-Hipparcos Cosmic Candles, ed. A. Heck 
   \& F. Caputo (Dordrecht: Kluwer), 111
\reference{} Clem, J. L., VandenBerg, D. A., Grundahl, F., \& Bell, R. A. 2004, 
  \aj, 127, 1227
\reference{} Davidge, T. J., \& Courteau, S. 1999, \aj, 117, 1297
\reference{} Demarque, P., Zinn, R., Lee, Y.-W., \& Yi, S. 2000, \aj, 119, 1398
\reference{} Ferraro, F. R., Paltrinieri, B., Fusi Pecci, F., Rood, 
  R. T., \& Dorman, B. 1998, \apj, 500, 311
\reference{} Girardi, L., Bertelli, G., Bressan, A., Chiosi, C., 
  Groenewegen, M. A. T., Marigo, P., Salasnich, B., \& Weiss, A. 2002, 
  \aap, 391, 195
\reference{} Isobe, T., Feigelson, E. D., Akritas, M. G., \& Babu, G. J. 
  1990, \apj, 364, 104
\reference{} Leavitt, H. S. 1912, Harvard Circular, 173 (reported by 
  E. C. Pickering) 
\reference{} Longmore, A. J., Fernley, J. A., \& Jameson, R. F. 1986, 
  \mnras, 220, 279 
\reference{} Pritzl, B. J., Smith, H. A., Catelan, M., \& Sweigart, A. V. 2002, 
  \aj, 124, 949 
\reference{} Salaris, M., Chieffi, A., \& Straniero, O. 1993, \apj, 414, 580 
\reference{} Stellingwerf, R. F. 1984, \apj, 277, 322
\reference{} Sweigart, A. V., \& Catelan, M. 1998, \apjl, 501, L63
\reference{} Tanvir, N. R. 1999, in Post-Hipparcos Cosmic Candles, ed. 
  A. Heck \& F. Caputo (Kluwer: Dordrecht), 17 
\reference{} van Albada, T. S., \& Baker, N. 1971, \apj, 169, 311
\reference{} van Albada, T. S., \& Baker, N. 1973, \apj, 185, 477
\reference{} VandenBerg, D. A., Swenson, F. J., Rogers, F. J., Iglesias, C. A., 
  \& Alexander, D. R. 2000, \apj, 532, 430 
\end{references}
\end{document}